\pdfoutput=1
\documentclass[conference,compsoc]{IEEEtran}

\usepackage[T1]{fontenc}
\usepackage{graphicx}
\usepackage{amsmath,amssymb,amsfonts}
\usepackage{algorithmic}
\usepackage{textcomp}
\usepackage{xcolor}
\usepackage{booktabs,array}
\usepackage{pbalance}
\usepackage{multirow}
\usepackage[hyphens]{url}
\usepackage{hyperref}
\usepackage{listings}
\usepackage{tikz}
\usetikzlibrary{arrows.meta,fit}
\usepackage{lucide-icons}
\usepackage{xspace}
\hypersetup{hidelinks, pdftitle={Towards Reliable Local Security Agents: Verifiable Post-Training for Linux Privilege Escalation}, pdfauthor={Philipp Normann, Andreas Happe, J\"urgen Cito, Daniel Arp}}
\usepackage[nolist,nohyperlinks]{acronym}
\usepackage[skins]{tcolorbox}
\usepackage{needspace}

\newcommand{\scenario}[1]{\textit{#1}}
\newcommand{\modelbackbone}{\textit{Qwen3 4B}\xspace}
\newcommand{\modelsft}{\textit{Qwen3 4B SFT}\xspace}
\newcommand{\modelrl}{\textit{PrivEsc-LLM 4B}\xspace}
\newcommand{\modelgemma}{\textit{Gemma~4~31B}\xspace}
\newcommand{\modelclaude}{\textit{Claude Opus 4.7}\xspace}
\newcommand{\modeldeepseek}{\textit{DeepSeek~V3.2}\xspace}
\newcommand{\modeldeepseekteacher}{\textit{DeepSeek V4 Flash}\xspace}
\newcommand{\chainreactor}{\textit{ChainReactor}\xspace}
\DeclareRobustCommand{\rewardvariant}[1]{{\normalfont\mbox{\textsc{#1}}}}
\definecolor{tracebg}{RGB}{248,249,251}
\definecolor{tracerule}{RGB}{208,214,222}
\definecolor{traceheader}{RGB}{31,41,55}
\definecolor{assistantbg}{RGB}{219,234,254}
\definecolor{toolcallbg}{RGB}{220,252,231}
\definecolor{toolresultbg}{RGB}{241,245,249}
\definecolor{promptbg}{RGB}{254,243,199}
\definecolor{toolcalltext}{RGB}{20,83,45}
\definecolor{toolcallborder}{RGB}{34,197,94}
\lstdefinestyle{traceexcerpt}{
  basicstyle=\tiny\ttfamily,
  backgroundcolor=\color{tracebg},
  frame=none,
  framesep=6pt,
  xleftmargin=6pt,
  xrightmargin=6pt,
  aboveskip=0pt,
  belowskip=0pt,
  breaklines=true,
  breakatwhitespace=false,
  columns=fullflexible,
}
\lstdefinestyle{toolcallexcerpt}{
  basicstyle=\tiny\ttfamily\color{toolcalltext},
  backgroundcolor=\color{white},
  rulecolor=\color{toolcallborder},
  frame=leftline,
  framesep=6pt,
  xleftmargin=8pt,
  xrightmargin=0pt,
  aboveskip=0pt,
  belowskip=0pt,
  breaklines=true,
  breakatwhitespace=false,
  columns=fullflexible,
}

\definecolor{findingfill}{RGB}{241,245,249}
\newtcolorbox{findingbox}[1]{%
  enhanced,
  colback=findingfill,
  colframe=traceheader,
  boxrule=0pt,
  leftrule=2.5pt,
  arc=0pt,
  outer arc=0pt,
  left=6pt,right=6pt,top=5pt,bottom=5pt,
  before skip=8pt,after skip=8pt,
  coltitle=traceheader,
  fonttitle=\footnotesize\bfseries,
  title={#1},
  attach title to upper,
  after title={\hspace{0.45em}},
  fontupper=\footnotesize,
}

\begin{acronym}
    \acro{LLM}{large language model}
    \acro{RL}{reinforcement learning}
    \acro{SFT}{supervised fine-tuning}
    \acro{CTF}{capture-the-flag}
    \acro{RLVR}{reinforcement learning with verifiable rewards}
    \acro{RLHF}{reinforcement learning from human feedback}
    \acro{SLM}{small language model}
    \acro{MoE}{mixture-of-experts}
    \acro{LoRA}{low-rank adaptation}
    \acro{QLoRA}{quantized low-rank adaptation}
    \acro{PPO}{proximal policy optimization}
    \acro{GRPO}{group relative policy optimization}
    \acro{AdamW}{Adam with decoupled weight decay}
    \acro{KL}{Kullback--Leibler}
\end{acronym}

\usepackage{eso-pic}

\AddToHookNext{shipout/foreground}{%
  \put(\LenToUnit{0.5\paperwidth},\LenToUnit{-0.8in}){%
    \makebox[0pt]{\normalsize Accepted at the 42nd Annual Computer Security Applications Conference (ACSAC 2026)}}%
  \put(\LenToUnit{0.5\paperwidth},\LenToUnit{-\paperheight+0.4in}){%
    \makebox[0pt][c]{\raisebox{0pt}[\height][0pt]{\fbox{\parbox{\dimexpr\textwidth-2\fboxsep-2\fboxrule\relax}{\footnotesize
      \textcopyright~2026 IEEE. Personal use of this material is permitted. Permission from IEEE must be obtained for all other uses, in any current or future media, including reprinting/republishing this material for advertising or promotional purposes, creating new collective works, for resale or redistribution to servers or lists, or reuse of any copyrighted component of this work in other works.}}}}}}
\begin{document}

\title{Towards Reliable Local Security Agents:\\
Verifiable Post-Training for Linux Privilege Escalation}

\author{%
  \IEEEauthorblockN{Philipp Normann, Andreas Happe, J\"urgen Cito, Daniel Arp}
  \IEEEauthorblockA{TU Wien, Vienna, Austria \\
  \{philipp.normann, andreas.happe, juergen.cito, daniel.arp\}@tuwien.ac.at}
}

\maketitle

\begin{abstract}

\acs{LLM} agents are increasingly important in security, yet leading systems are typically closed-source and cloud-hosted, limiting reproducibility and their use with sensitive data.
This creates a need for small, local models that can perform security tasks under strict resource constraints, though effective methods for developing them remain underexplored.

In this paper, we address this gap by proposing a two-stage post-training recipe that turns a small local language model into a security agent.
To this end, we focus on Linux privilege escalation as a representative setting to systematically study the training of local models, as the task is automatically verifiable and requires multi-step interactive reasoning.
Using an experimental setup that mitigates benchmark leakage, we post-train a small 4B model in two stages: \acl{SFT} on traces from procedural privilege-escalation environments, followed by \acl{RLVR}.
On a held-out benchmark of 12 Linux privilege-escalation scenarios, \acl{SFT} nearly doubles the baseline success rate under a tight budget of 20 interaction rounds, and subsequent \acl{RL} improves our model, \modelrl, to 93.3\% success, ahead of all other evaluated models that fit a single consumer GPU.
At the same time, its expected inference cost per successful escalation is more than 80$\times$ lower than that of \modelclaude.

Our findings show that small local models can be adapted to complex security tasks.
We document the challenges involved and offer guidance for transferring this recipe to other settings.

\end{abstract}


\section{Introduction}
\label{sec:introduction}

\acs{LLM}-based systems are demonstrating practical vulnerability discovery capabilities, as highlighted by AIxCC~\cite{zhangAIxCCSoK26}, CyberGym~\cite{wangCyberGymICLR26}, and prior work on \acs{LLM}-assisted fuzzing~\cite{xiaFuzz4AllICSE24}. Mozilla reports an order-of-magnitude rise in monthly Firefox security fixes, from 20--30 per month in 2025 to 423 in April~2026, most of them surfaced by agentic \acs{LLM} pipelines~\cite{mozillaHardeningFirefox2026}. As these capabilities mature, the limiting factor shifts from whether {\acs{LLM}}s can find vulnerabilities to whether the resulting systems are reproducible, budget-efficient, and deployable without sending sensitive system state to external infrastructure.

The strongest systems today are mostly closed stacks: closed-source code, proprietary models, and cloud-only inference. That is a deployment blocker whenever security-relevant host data cannot leave local trust boundaries: sending it to a third-party API exposes it to provider-side logging and retention, data leakage, and, if the data is later used for training, extraction attacks~\cite{carliniExtractingTrainingData2021}. Post-training a small open-weight model on the task is one response, but the design choices that govern whether it becomes a reliable agent have not been systematically isolated for small security agents. These choices include how traces are collected, how much reasoning is retained, and which reward terms are used. This motivates our question: \emph{what post-training recipe makes a small, locally deployable \acs{LLM} agent reliable on a verifiable security task, and how does it compare to frontier models?}

We answer this with a two-stage recipe. \Ac{SFT} on filtered teacher traces first teaches successful reconnaissance-to-root behavior, and \ac{RLVR} on verified root access then trains the policy to act efficiently on its own observations. Applied to a 4B open-weight backbone, the recipe produces \modelrl. On a held-out static benchmark, success within 20 interaction rounds rises from 40.8\% for the base model to 79.2\% after \ac{SFT} and 93.3\% after \ac{RL}, while \modelclaude solves every run. With only five rounds, \modelrl outperforms Claude (78.3\% versus 66.7\%). It is the strongest evaluated model that fits a single consumer GPU, ahead of the nearly eight times larger \modelgemma (80.8\%), and costs an estimated 83$\times$ less per success than Claude. Post-training can therefore bring a small local model close to frontier reliability on a verifiable security task.

Linux privilege escalation provides a controlled testbed for this recipe. Starting from a low-privileged account, the agent must enumerate an unfamiliar host, identify an exploitable configuration, and execute the commands that obtain root. The benchmark covers the vectors taught in standard Linux privilege-escalation training and checked by practitioners on a newly compromised host: credential discovery and reuse, SUID and sudo abuse, scheduled-task manipulation, and container escape. Two professional penetration testers judged the selection representative~\cite{happeKaplanCitoEMSE26}. Each scenario requires reconnaissance and multi-step exploitation, and a root-owned canary file verifies success automatically.

Training uses procedural environments that exclude the benchmark's binaries, credentials, paths, and solution fragments, so gains reflect transferable behavior rather than memorized solve paths. Ten of the 12 benchmark scenarios belong to vulnerability families seen in training, but none of their concrete instances do. For example, the training SUID allowlist covers 24 abusable Unix binaries but excludes the benchmark's \texttt{find} and Python binaries, so the agent must identify an unseen binary and construct its specific escape. \scenario{Docker group escape} is withheld as an entire family and serves as a cross-family stress test. Unlike previous single-run or best-of-few benchmarks~\cite{wangCyberGymICLR26,happeKaplanCitoEMSE26,zhangCyBenchICLR25}, we measure success at fixed round budgets over 10 repeated runs per scenario with confidence intervals.

To identify which design choices carry these gains, we ablate trace collection and reasoning format for \ac{SFT} and compare reward terms for \ac{RL}, selecting on procedural holdout before evaluating the final model on the static benchmark. We release code, configurations, datasets, models, and evaluation traces through the artifact repository.\footnote{\label{fn:repo}\begin{tabular}[t]{@{}l@{\ }l@{}}GitHub: & \url{https://github.com/sailab-vienna/privesc-llm}\\Zenodo: & \url{https://doi.org/10.5281/zenodo.23556797}\end{tabular}}

We make four contributions:
\begin{enumerate}
\item \textbf{Leakage-controlled generalization tests.} We generate training environments that exclude all benchmark-specific values and audit every trace for leakage, so evaluation measures transfer rather than memorization. A withheld vulnerability family additionally probes cross-family transfer.
\item \textbf{Systematic post-training recipe ablation.} We isolate the contribution of each design choice in our post-training pipeline. The most reliable local agent comes from \ac{SFT} on teacher traces without hidden solution data but with full reasoning, followed by \ac{RL} with a latency-based cost bonus.
\item \textbf{Budgeted repeated-run reliability.}
We run each model 10 times per scenario under fixed round budgets with confidence intervals. \modelrl reaches 93.3\% success within 20 rounds, failing only on the withheld Docker family.
\item \textbf{Training and inference cost accounting.}
We estimate post-training cost and the expected inference cost per successful root, including failed attempts. At $r{=}20$, \modelrl is about 83$\times$ cheaper than \modelclaude and amortizes this training cost after roughly 700 successful escalations.
\end{enumerate}

\section{Background}
\label{sec:background}

Before presenting our method, we briefly recap the Linux privilege-escalation task we study, the post-training and reinforcement learning paradigms we apply, the agent harness design, and the role of procedural generation in leakage-controlled evaluation.

\textbf{Post-exploitation Linux privilege escalation.}
We study the phase after an attacker already has a foothold on a Linux host, not the initial external compromise. The starting point is a low-privileged account or shell inside the system. From there, the task is to reach \texttt{root} by finding and exploiting local weaknesses such as unsafe \texttt{sudo} configurations, credential reuse, writable files executed by privileged services, or binaries whose built-in features can be repurposed for escalation, as catalogued in GTFOBins, a community-curated index of Unix binaries that can be abused~\cite{happeKaplanCitoEMSE26,gtfobins}.

\textbf{Pretraining and post-training.}
Modern language models are first pretrained on large corpora, where they acquire broad linguistic and factual knowledge~\cite{brownGPT3,petroniLMKB,robertsParametricKnowledge,changFactualKnowledge}, and then post-trained for downstream use through instruction tuning, preference optimization, or task-specific adaptation~\cite{chungFlan,ouyangInstructGPT}. We apply our post-training pipeline to an instruction-tuned Qwen3~\cite{qwen3} backbone.

\textbf{RLHF and RLVR.}
In \ac{RL}, a policy improves by maximizing expected return over trajectories of observations, actions, and rewards~\cite{suttonBarto}; for language agents, observations are the running context plus tool outputs, and actions are generated tokens and tool calls. \Ac{RLHF} optimizes a policy against a learned reward model from human preferences with a \ac{KL} drift penalty~\cite{ouyangInstructGPT,christianoPreferences}.
\ac{RLVR} replaces the learned reward with automatic verification and has been effective on math and coding tasks~\cite{deepseekmath,deepseekR1Nature,wenRLVRICLR26}. Because \texttt{root} access is equally binary and verifiable directly by the environment, and multi-step agent training suffers from distribution shift under pure imitation~\cite{rossDAgger}, \ac{RLVR} is a natural fit for privilege escalation.

\textbf{Agent harness.}
An agent harness is the runtime layer that turns a language model into an executable agent. It provides the prompt context, exposes tools, executes commands, returns observations, resets the environment between runs, and records the full interaction trace. In tool-use settings, the harness effectively defines the action space and the form of feedback. If it changes between data collection and evaluation, a failure may reflect an interface mismatch rather than a weakness of the policy~\cite{react,toolformer,terminalBench}.

\textbf{Procedural generation.}
Procedural generation creates many fresh task instances from templates or seeds instead of reusing one fixed environment~\cite{cobbeProcgen}. In \ac{RL}, this is a standard way to train on a task distribution and test whether a policy transfers to held-out variants rather than memorizing one layout~\cite{cobbeProcgen,cobbeCoinRun}. In security, that can mean varying usernames, passwords, service names, file paths, or vulnerable artifacts while preserving the same exploit family. This helps distinguish brittle pattern matching from policies that generalize across new instances~\cite{arpDosDonts,evertzChasingShadows}.

\begin{figure*}[t]
  \centering
  \makebox[\linewidth][c]{%
    \resizebox{\linewidth}{!}{%
      \begin{tikzpicture}[x=1.3cm,y=1cm,line join=round,line cap=round]
\definecolor{archblue}{RGB}{76,120,168}
\definecolor{archbluefill}{RGB}{244,247,252}
\definecolor{archbluebar}{RGB}{227,236,247}
\definecolor{archgreen}{RGB}{89,161,106}
\definecolor{archgreenfill}{RGB}{244,249,245}
\definecolor{archgreenbar}{RGB}{227,239,230}
\definecolor{archorange}{RGB}{217,147,74}
\definecolor{archorangefill}{RGB}{253,247,240}
\definecolor{archorangebar}{RGB}{247,232,214}
\definecolor{archred}{RGB}{191,78,78}
\definecolor{archredfill}{RGB}{253,243,243}
\definecolor{archgray}{RGB}{127,133,141}
\definecolor{archgrayfill}{RGB}{247,248,250}
\definecolor{archgraybar}{RGB}{233,236,240}
\definecolor{archink}{RGB}{42,47,56}

\newcommand{\archicon}[1]{\raisebox{-0.08em}{\lucideicon[height=0.80em]{#1}}}

\tikzset{
  flow/.style={-{Latex[length=2.4mm,width=1.6mm]}, line width=0.95pt, draw=archink!92},
  card/.style={rounded corners=5pt, line width=0.9pt},
  band/.style={rounded corners=5pt, line width=0.8pt},
  title/.style={font=\sffamily\bfseries\footnotesize, text=archink, align=center},
  item/.style={rounded corners=3pt, line width=0.65pt, fill=white, inner xsep=4pt, inner ysep=3pt, font=\sffamily\scriptsize, text=archink, align=center},
  badge/.style={rounded corners=3pt, line width=0.85pt, draw=archred, fill=archredfill, inner xsep=6pt, inner ysep=2.75pt, font=\sffamily\bfseries\scriptsize, text=archred, align=center},
}

\filldraw[card, draw=archblue, fill=archbluefill] (0,0) rectangle (2.5,4.5);
\filldraw[band, draw=archblue, fill=archbluebar] (0,3.5) rectangle (2.5,4.5);
\node[title] at (1.25,4) {Procedural Training\\Environments};
\node[item, draw=archblue, text width=1.75cm] at (1.25,2.80) {10 generator\\families};
\node[item, draw=archblue, text width=1.75cm] at (1.25,1.75) {randomized creds,\\paths, configs};
\node[item, draw=archred, text width=1.75cm] at (1.25,0.70) {\archicon{shield-ban}\, benchmark\\exclusions};
\node[inner sep=0pt, outer sep=0pt, fit={(0,0) (2.5,4.5)}] (proc_panel) {};

\filldraw[card, draw=archblue, fill=archbluefill] (3,0) rectangle (5.5,4.5);
\filldraw[band, draw=archblue, fill=archbluebar] (3,3.5) rectangle (5.5,4.5);
\node[title] at (4.25,4) {Trace\\Collection};
\node[item, draw=archblue, text width=1.75cm] at (4.25,2.85) {\archicon{bot}\, open-weight\\teacher model};
\node[item, draw=archblue, text width=1.75cm] at (4.25,1.90) {2{,}000 teacher\\traces};
\node[item, draw=archgreen, text width=1.75cm] at (4.25,0.80) {\archicon{funnel}\, filtered successful\\traces};
\node[inner sep=0pt, outer sep=0pt, fit={(3,0) (5.5,4.5)}] (trace_panel) {};

\filldraw[card, draw=archgreen, fill=archgreenfill] (6,0) rectangle (8,4.5);
\filldraw[band, draw=archgreen, fill=archgreenbar] (6,3.5) rectangle (8,4.5);
\node[title] at (7,4) {Stage 1: SFT};
\node[item, draw=archgreen, text width=1.8cm, font=\sffamily\scriptsize] (s1a) at (7,2.5) {Qwen3 4B\\+ LoRA};
\node[item, draw=archgreen, text width=1.8cm, font=\sffamily\scriptsize\bfseries] (s1b) at (7,1.0) {Qwen3 4B\\SFT};
\draw[flow] (s1a.south) -- (s1b.north);
\node[inner sep=0pt, outer sep=0pt, fit={(6,0) (8,4.5)}] (sft_panel) {};

\filldraw[card, draw=archorange, fill=archorangefill] (8.5,0) rectangle (13.0,4.5);
\filldraw[band, draw=archorange, fill=archorangebar] (8.5,3.5) rectangle (13.0,4.5);
\node[title] at (10.75,4) {Stage 2: RLVR};
\node[item, draw=archorange, text width=2.25cm, font=\sffamily\scriptsize\bfseries] (rl_policy) at (10.75,3.00) {\archicon{cpu}\, PrivEsc-LLM 4B};
\node[item, draw=archblue, text width=1.50cm] (rl_env) at (12.08,1.90) {\archicon{terminal}\, rollouts in\\procedural\\envs};
\node[item, draw=archgreen, text width=2.00cm] (rl_reward) at (10.75,0.65) {\archicon{shield-check}\, verified root\\\archicon{timer}\, measured cost};
\node[item, draw=archorange, text width=1.20cm] (rl_update) at (9.42,1.90) {\archicon{refresh-cw}\, RLVR policy\\update};
\draw[flow] (rl_policy.east) to[out=0,in=90, looseness=1.55] (rl_env.north);
\draw[flow] (rl_env.south) to[out=-90,in=0, looseness=1.55] (rl_reward.east);
\draw[flow] (rl_reward.west) to[out=180,in=-90, looseness=1.55] (rl_update.south);
\draw[flow] (rl_update.north) to[out=90,in=180, looseness=1.55] (rl_policy.west);
\node[inner sep=0pt, outer sep=0pt, fit={(8.5,0) (13.0,4.5)}] (rlvr_panel) {};

\node[badge] (badge) at (13.42,4.95) {\archicon{lock}\, leakage control};
\draw[draw=archred, fill=archred!7, line width=1.0pt, rounded corners=2pt] (13.32,0) rectangle (13.52,4.5);
\draw[draw=archred, line width=0.95pt] (13.42,0.2) -- (13.42,4.3);
\node[inner sep=0pt, outer sep=0pt, fit={(13.32,0) (13.52,4.5)}] (leakage_barrier) {};

\filldraw[card, draw=archgray, fill=archgrayfill] (13.82,0) rectangle (16.32,4.5);
\filldraw[band, draw=archgray, fill=archgraybar] (13.82,3.5) rectangle (16.32,4.5);
\node[title] at (15.07,4) {Held-Out\\Evaluation};
\node[item, draw=archgray, text width=1.75cm] at (15.07,2.85) {12 static\\scenarios};
\node[item, draw=archgray, text width=1.75cm] at (15.07,1.90) {10 runs per\\scenario};
\node[item, draw=archgray, text width=1.75cm] at (15.07,0.80) {\archicon{timer}\, fixed budgets\\+ success/cost};
\node[inner sep=0pt, outer sep=0pt, fit={(13.82,0) (16.32,4.5)}] (eval_panel) {};

\draw[flow] (proc_panel.east) -- (trace_panel.west);
\draw[flow] (trace_panel.east) -- (sft_panel.west);
\draw[flow] (sft_panel.east) -- (rlvr_panel.west);
\draw[flow] (rlvr_panel.east) -- (leakage_barrier.west);
\draw[flow] (leakage_barrier.east) -- (eval_panel.west);

\pgfresetboundingbox
\path[use as bounding box] ([shift={(-0.5pt,-0.5pt)}]proc_panel.south west) rectangle ([shift={(0.5pt,0.5pt)}]eval_panel.east |- badge.north);
\end{tikzpicture}%
%
    }%
  }
  \caption{Leakage-controlled training and evaluation pipeline. Training uses only procedurally generated privilege-escalation environments with benchmark exclusions. Successful traces from an open-weight teacher are filtered and used for Stage~1 \acs{SFT} on \modelbackbone with \acs{LoRA}. Stage~2 \acs{RLVR} trains against procedural environments with verifiable root-success and efficiency-shaping rewards, and produces \modelrl. The held-out static benchmark is used only for repeated fixed-budget evaluation.}
  \label{fig:architecture_overview}
\end{figure*}
\section{Methodology}
\label{sec:methodology}

Figure~\ref{fig:architecture_overview} summarizes our pipeline. We first define the agent's task and interface (Section~\ref{sec:benchmark}), then describe the procedural training environments and their separation from the static benchmark (Section~\ref{sec:procedural}), \acs{SFT} (Section~\ref{sec:agent}), and \acs{RL} with its verifiable reward (Section~\ref{sec:training}).

\subsection{Task and Agent Interface}
\label{sec:benchmark}

We study post-exploitation Linux privilege escalation as a controlled interactive decision problem: starting from SSH access as a low-privileged user, the agent must reach \emph{root}, our binary success signal. Success is automatically checked by reading a per-run random secret from a root-owned canary file. Many security tasks share this kind of cheap, automatic verification, which makes them well suited to \ac{RLVR} without human grading or a learned reward model.

We use native tool calling with two tool interfaces: \texttt{exec\_command} for arbitrary shell commands and \texttt{test\_credentials} for checking discovered credentials. We define a \emph{round} as a single \acs{LLM} call together with the execution of any tool calls it issues.
Multiple tool calls in the same response are dispatched in parallel and counted as one round. A \emph{run} is one full trajectory in a fresh container, starting from the initial low-privileged SSH session and terminating either when root is verified or when a fixed per-run round cap is reached. If a response contains no tool call, the harness injects a short nudge so the agent cannot stall the loop. The same contract is used in training, validation, and evaluation, so behavioral differences reflect the policy itself and not the agent harness.

Final performance is reported on the 12-scenario held-out \emph{Linux PrivEsc} benchmark~\cite{happeKaplanCitoEMSE26}. Let $H_{\mathrm{root}}$ be the first round in which a run achieves root, with $H_{\mathrm{root}}{=}\infty$ if root is not reached within the horizon. The budgeted success probability at budget $r$ is $P(H_{\mathrm{root}}\le r)$, estimated across $N$ runs as $\tfrac{1}{N}\sum_{i=1}^{N}\mathbf{1}\{H_i\le r\}$. We refer to this as success within $r$ rounds. Low budgets matter operationally: each additional round is another \acs{LLM} inference call, and inference dominates agent-loop cost, so policies that achieve root early are cheaper and faster to run.

\subsection{Procedural Environments and Holdout Design}
\label{sec:procedural}

To prevent the static evaluation benchmark from leaking into training data (\emph{benchmark leakage})~\cite{dengDataContamination2024}, we train exclusively on procedural scenarios and reserve the static benchmark for evaluation only. Using procedurally varied environments as the training distribution is consistent with prior work that treats procedural generation as a direct testbed for generalization in \ac{RL}~\cite{cobbeProcgen}. Our distribution contains 10 generator families chosen to cover most of the same privilege-escalation classes as the static benchmark while allowing controlled variation. Each procedural instance is generated from a deterministic seed and includes (i)~a setup script executed after container start, (ii)~scenario metadata, and (iii)~reference exploit traces used for automatic exploitability checks and trace-collection guidance. We randomize all scenario-specific strings that are realistically variable, such as usernames, passwords, key names, file locations, service identifiers, and file contents, while keeping structural artifacts like canonical binary paths unchanged. This design makes literal benchmark-value memorization insufficient and tests whether the policy transfers across value-disjoint procedural instances within shared vulnerability families. Table~\ref{tab:generator_catalog} in \hyperref[app:repro_details]{Appendix~A} summarizes all 10 generators and the weakness each one plants.

\subsubsection{Holdout Design}
\label{sec:holdout}
We use a procedural holdout as the internal validation oracle and reserve the static benchmark exclusively for final evaluation. All training-time decisions, including reward variant, trace format, checkpoint, and hyperparameters, are based only on procedural data (Table~\ref{tab:training_selection_protocol}), and the deployment prompt is fixed beforehand; the static benchmark is never consulted during training. To prevent \emph{holdout leakage}, training and holdout use \emph{value-disjoint generator profiles}: the two profiles draw from entirely separate parameter pools across all generator families, confirmed at zero overlap by a post-hoc audit. The model therefore encounters different GTFOBins binaries, passwords, usernames, filenames, and cron and SSH configurations at validation time than during training. For example, the training SUID allowlist contains 24 binaries while the holdout uses 3 disjoint ones, and credential-disclosure scenarios likewise use 24 versus 3 disjoint credential-file paths.

To prevent benchmark leakage, generator-level rules remove benchmark-specific values, such as binaries, credentials, and file paths, from the generator sampling spaces where applicable, which keeps these values out of the training data. Table~\ref{tab:scenario_taxonomy} in \hyperref[app:repro_details]{Appendix~A} lists these values for each benchmark scenario. Every assembled trace is scanned against the same rule set and rejected on any match. The audit reports zero matches in the procedural pools. The 12 benchmark scenarios partition into three distribution-status classes. Ten scenarios are \emph{covered} by a procedural generator family, with all benchmark-specific values excluded. \scenario{Sudo no password} grants passwordless sudo for all commands and has no direct procedural analog, since the separate \scenario{Sudo GTFOBins} family covers only restricted passwordless sudo for a single binary. \scenario{Docker group escape} is a \emph{withheld family} and is deliberately omitted from both procedural profiles. We therefore treat its success rate as a cross-family stress test rather than evidence of in-family generalization.

\subsection{Supervised Fine-Tuning}
\label{sec:agent}

\ac{SFT} prepares the base model for \ac{RL} by giving it both privilege-escalation task knowledge and the ability to act within our harness through valid \texttt{exec\_command} and \texttt{test\_credentials} calls, so that it solves a non-trivial fraction of procedural runs out of the box. This matters because verifier-based \ac{RL} can only reinforce trajectories that occasionally succeed. Cold-start supervised demonstrations are the standard remedy, used both in instruction tuning~\cite{ouyangInstructGPT} and in recent \ac{RLVR} systems~\cite{deepseekR1Nature}.

\subsubsection{Trace Collection}
We collect supervised traces in procedural environments using \modeldeepseekteacher~\cite{deepseekV4FlashModelCard} accessed through OpenRouter (the exact model identifier, provider routing, and per-regime sampling parameters are reported in \hyperref[app:teacher_inference_config]{Appendix~A}). We chose this teacher because it is open-weight, offers a favorable price-to-performance trade-off, and can solve our procedural environments in both guided and unguided modes within a small number of retries. Together, these properties make large-scale trace collection across all 10 generator families practical. We treat trace design as an experimental variable along two axes.

The first is \emph{teacher information}: the teacher either solves the scenario from the same observable state as the deployed student (\emph{unguided}) or sees hidden reference-solution data that is removed before training (\emph{guided}). Guided traces succeed more often and are shorter, which gives cleaner targets, but they may contain steps the student cannot infer from observable state alone. Unguided traces are noisier but use only what the student observes at deployment.

The second axis is the \emph{amount of retained reasoning} in the training target: none, a concise rewrite, or the original long teacher reasoning. More reasoning may transfer richer planning structure, but it also lengthens sequences and may exceed what a small student model can absorb given its limited capacity. We construct all six combinations and select among them on procedural holdout alone (Section~\ref{sec:trace_ablation}); guided prompt details are in \hyperref[app:prompt_templates]{Appendix~C}, and all training and evaluation prompts use the deployment-time prompt, never the hidden solution data.

\subsubsection{SFT Assembly}
The three reasoning variants are derived from the same collected traces: the \emph{long-reasoning} variant uses the original teacher reasoning unchanged, the \emph{no-reasoning} variant strips reasoning while preserving the tool-call structure, and the \emph{short-reasoning} variant uses the same teacher model to rewrite each assistant message into a strictly shorter reasoning step. For each of the resulting six trace-design combinations, we assemble balanced splits of 2{,}000 training and 200 validation traces, derived from 200 and 20 samples per generator. Before balancing, we apply quality filters: successful runs only, non-empty reasoning, rejection of \emph{solution leakage} (assistant text that refers to the hidden solution data, found by an automatic phrase scan), and the benchmark-exclusion scan from Section~\ref{sec:holdout} applied per trace (benchmark-leakage guard). Post-hoc audits report zero solution- or benchmark-leakage matches in the assembled splits. During dataset assembly, the trace-collection system prompt is replaced with the deployment-time system prompt used for training and evaluation.

\subsubsection{Stage 1 Training}
We fine-tune \modelbackbone with \ac{LoRA} adapters~\cite{lora} using Unsloth~\cite{unsloth} on the assembled training split. Checkpoint and hyperparameter selection uses only the procedural validation split, so the static benchmark cannot influence these choices. Hyperparameter search follows a fixed protocol on procedural data only: a log-spaced learning-rate sweep, a \ac{LoRA}-rank sweep informed by prior findings~\cite{lorawr}, and multi-seed confirmation of the selected hyperparameters.

\subsection{Reinforcement Learning}
\label{sec:training}

A reliable security agent has to act well under its own observations, recovering from dead ends, exploring new attack paths, and reaching root within bounded interaction. \ac{SFT} alone does not guarantee this: imitation of stronger teachers narrows the apparent style gap without closing the underlying capability gap~\cite{gudibandeFalsePromise2024}, and training on offline teacher demonstrations leaves the policy exposed to distribution shift whenever it visits states the teacher never showed~\cite{rossDAgger}. Stage~2 addresses this gap by training under the policy's own induced state distribution, using verified root as the primary signal alongside optional efficiency shaping.

\subsubsection{Reward Design}
We seek the minimal verifier-aligned reward that reflects the operational goal of an interactive security agent without injecting task-specific privilege-escalation heuristics. Because auxiliary terms can shift the optimal policy when they are misaligned with the underlying objective~\cite{ngRewardShaping}, we evaluate each term in isolation before combining them (Equation~\ref{eq:reward_variants}).

\emph{Outcome term.}
Binary verified root success is the cleanest \ac{RLVR} signal, consistent with recent \ac{RLVR} systems that rely on rule-based verifiers when correctness can be checked automatically~\cite{deepseekR1Nature}. For an episode $e$,
\begin{equation}
R_{\text{out}}(e) = 2\cdot\mathbf{1}\!\{\mathrm{root}(e)\} - 1.
\end{equation}
In a long-horizon tool-use setting, however, verified success alone under-specifies efficiency and admits wasteful successful trajectories. We therefore consider two complementary efficiency signals, each as an additive shaping term applied only on successful episodes.

\emph{Round shaping.}
The round term directly tracks the budgeted evaluation metric $P(H_{\mathrm{root}}\le r)$ and follows prior decompositions of \acs{LLM}-agent efficiency into trajectory-level step efficiency and per-step cost~\cite{chenDEPOAAAI26}:
\begin{equation}
R_{\text{round}}(e) = \mathbf{1}\!\{\mathrm{root}(e)\}\operatorname{clip}_{[0,1]}\!\left(1-\tfrac{H_{\mathrm{root}}(e)}{h_{\max}}\right),
\end{equation}
where $h_{\max}$ is the training round budget. Round count is only a proxy for interaction cost, since one assistant round may issue multiple parallel tool calls. Optimizing the proxy can, in principle, diverge from the underlying objective, a known reward-hacking failure mode~\cite{skalseRewardHackingNeurIPS22}. We use $R_{\text{round}}$ at unit coefficient so the \rewardvariant{Outcome{+}Round} variant remains interpretable as a surrogate for the round-budgeted metric.

\emph{Cost shaping.}
As an alternative efficiency signal, the cost term prices model inference and executed tool runtime:
\begin{equation}
R_{\text{cost}}(e) = \mathbf{1}\!\{\mathrm{root}(e)\}\operatorname{clip}_{[0,1]}\!\left(1-\tfrac{C(e)}{C_{\text{ref}}}\right),
\end{equation}
with episode cost
\begin{equation}
\label{eq:cost_term}
C(e) = \sum_{t=1}^{\tau(e)} \hat{\ell}_t^{\text{LLM}} + \sum_{t=1}^{\tau(e)}\sum_{k=1}^{m_t} \hat{\ell}_{t,k}^{\text{tool}},
\end{equation}
where $\tau(e)$ is the number of rounds in episode $e$, $m_t$ is the number of executed tool calls in round $t$, and per-call latencies are clipped per modality, $\hat{\ell}_t^{\text{LLM}} = \min\{\ell_t^{\text{LLM}}, c_{\text{clip}}^{\text{LLM}}\}$ and $\hat{\ell}_{t,k}^{\text{tool}} = \min\{\ell_{t,k}^{\text{tool}}, c_{\text{clip}}^{\text{tool}}\}$~(see Appendix~\ref{app:reward_cost_calibration} for calibration details). Measured latency is domain-agnostic, charges each parallel tool call separately, and unlike a tool-call count cannot be gamed by collapsing several shell operations into one open-ended \texttt{exec\_command}. The trade-off is that latency mixes policy behavior with infrastructure effects (server load, batching); the per-modality clip caps spikes, and rollouts of the same instance share infrastructure so relative comparisons stay roughly fair. The term is auxiliary: $R_{\text{cost}}$ is zero on failed episodes, clipped to $[0,1]$, and weighted by $\lambda_{\text{cost}}$, so cost shaping can rerank successful trajectories but cannot substitute for verified root success.

\emph{Interface guardrail.}
All reward variants include a small interface penalty for invalid or missing tool calls:
\begin{equation}
R_{\text{iface}}(e) = -\lambda_{\text{iface}}\,\mathbf{1}\!\{\mathrm{iface}(e)\}.
\end{equation}
The indicator fires at most once per episode when a non-terminal assistant response cannot be executed by the fixed harness: no executable tool call after parser normalization, malformed tool-call syntax, an unknown tool name, or schema-invalid arguments. Exact constants are reported in \hyperref[app:reward_details]{Appendix~A}. Small rule-based protocol rewards are common in \ac{RLVR} and tool-use \ac{RL} when the interaction contract is directly checkable~\cite{deepseekR1Nature,qianToolRLNeurIPS25}.

\emph{Reward variants.}
We compare four variants on procedural holdout, crossing round shaping and cost shaping in a $2{\times}2$ design over the shared $R_{\text{out}}$ core:
\begin{equation}
\label{eq:reward_variants}
\begin{aligned}
\text{\rewardvariant{Outcome}}:\;            &R_{\text{out}}(e), \\
\text{\rewardvariant{{+}Round}}:\;           &R_{\text{out}}(e) + R_{\text{round}}(e), \\
\text{\rewardvariant{{+}Cost}}:\;            &R_{\text{out}}(e) + \lambda_{\text{cost}}\,R_{\text{cost}}(e), \\
\text{\rewardvariant{{+}Round{+}Cost}}:\;    &R_{\text{out}}(e) + R_{\text{round}}(e) + \lambda_{\text{cost}}\,R_{\text{cost}}(e).
\end{aligned}
\end{equation}
The total reward is $R(e) = R_{\text{variant}}(e) + R_{\text{iface}}(e)$, with the same guardrail definition and coefficient in every variant. We fix $\lambda_{\text{cost}}{=}0.1$ and $\lambda_{\text{iface}}{=}0.05$ before \ac{RL}: cost contributes at most 0.1 and an interface violation subtracts 0.05 at most once per episode, so every success receives a higher reward than every failure. The coefficients were not swept; the ablation compares reward terms at these fixed values.

A process reward could improve credit assignment, but we lack a reliable family-independent verifier of intermediate progress, a challenge also reported for DeepSeek-R1~\cite{deepseekR1Nature}. We therefore combine step-level demonstrations from \ac{SFT} with verified end-to-end feedback during \ac{RL}.

\subsubsection{Stage 2 Training}
We initialize from the selected \ac{SFT} checkpoint and train with Prime-RL~\cite{primerl}, with environment feedback exposed through a custom Verifiers~\cite{verifiers} wrapper. The objective resembles \ac{GRPO}~\cite{deepseekmath} with four implementation differences relevant to our setting: (i)~fully asynchronous rollout and learner execution with bounded staleness; (ii)~no within-group reward-standard-deviation normalization; (iii)~episode-level outcome rewards from direct environment verification; and (iv)~a token loss with trainer-vs-inference importance ratios plus a probability-difference clip mask. Asynchrony amortizes the cost of multi-round tool-using rollouts at the price of controlled off-policyness, and procedural generators are sampled round-robin to stabilize per-generator coverage. Checkpoints are selected on procedural holdout (Section~\ref{sec:reward_ablation}); cost-term clip and reference constants are calibrated before \ac{RL} from a timing benchmark of the \ac{SFT} initialization, with values and procedure reported in \hyperref[app:reward_details]{Appendix~A}. The 20-round training horizon matches the primary evaluation budget, and evaluation continues to 60 rounds to measure later recovery.

\section{Experimental Setup}
\label{sec:setup}

\textbf{\acs{LLM} models and baselines.}
Table~\ref{tab:models} lists the primary models we evaluate, together with their exact model identifiers, deployment method, and evaluated context length, following the recommendation of Evertz et al.~\cite{evertzChasingShadows} to reduce model-version ambiguity in \acs{LLM} evaluations. \modelbackbone serves as the unmodified local backbone baseline, while \modelsft and \modelrl are our two specialized \acs{LoRA} adapters trained on top of this backbone and served locally with vLLM~\cite{kwonPagedAttention}. Beyond the Qwen3 family, we also compare against \modelgemma~\cite{gemma4ModelCard} as a larger local open-weight reference point, and against \modeldeepseek~\cite{deepseekV3_2} and \modelclaude~\cite{claudeOpus47SystemCard} as API-served frontier comparisons accessed through OpenRouter~\cite{openrouter2026pricing}. All three comparison models were selected based on Terminal-Bench~\cite{terminalBench} rankings and internal pilot evaluations. Per-model sampling parameters and context caps for each model family are reported in Table~\ref{tab:inference_config} of \hyperref[app:inference_config]{Appendix~A}.

Supporting evaluations add Qwen3 8B and 14B~\cite{qwen3}, Llama 3.2 3B Instruct~\cite{llama32ModelCard}, and the teacher \modeldeepseekteacher (Table~\ref{tab:supplementary_baselines}). They test whether larger Qwen3 models or a similarly sized model from another family can match our post-trained agent without task-specific training, and how close the student comes to its teacher.

\textbf{Non-\acs{LLM} baselines.}
We further compare against \chainreactor~\cite{dePasqualeChainReactorUSENIX24}, a state-of-the-art method for automated privilege-escalation chain discovery on Unix systems, and relate our results to two tool-based exploitation frameworks (Traitor~\cite{traitor}, pwncat-cs~\cite{pwncatcs}) and a professional penetration tester evaluated on the same benchmark in prior work~\cite{happeKaplanCitoEMSE26}.

\begin{table}[t]
  \centering
  \caption{Primary comparison models with their evaluated context caps (Ctx). Local open-weight models use \textnormal{vLLM}; Table~\ref{tab:inference_config} reports serving precision and inference settings.}
  \label{tab:models}
  \footnotesize\setlength{\tabcolsep}{3pt}
  \begin{tabular}{llll}
    \toprule
    Model & Exact ID & Deploy & Ctx \\
    \midrule
    \modelbackbone   & \texttt{Qwen3-4B-Instruct-2507}~\cite{qwen3FourBInstruct2507ModelCard}    & vLLM & 32K \\
    \modelsft        & base + \acs{LoRA} rank 8       & vLLM & 32K \\
    \modelrl         & base + \acs{LoRA} rank 8       & vLLM & 32K \\
    \modelgemma      & \texttt{google/gemma-4-31B-it}~\cite{gemma4ModelCard}     & vLLM & 32K \\
    \modeldeepseek   & \texttt{deepseek/deepseek-v3.2}~\cite{deepseekV3_2ModelCard}    & API & 128K \\
    \modelclaude     & \texttt{anthropic/claude-opus-4.7} & API & 200K \\
    \bottomrule
  \end{tabular}
\end{table}

\textbf{Training.}
We fine-tune \modelbackbone with \ac{QLoRA}~\cite{dettmersQLoRA} using Unsloth on 4$\times$H100 GPUs and the filtered procedural training split described in Section~\ref{sec:agent}. We perform the reinforcement learning step with Prime-RL on 4$\times$H100 GPUs, initialized from the \ac{SFT} checkpoint. We sample procedural generators in round-robin mode with a 20-round training horizon. We train for 1{,}000 steps with a batch size of 80 and 8 rollouts per instance. Table~\ref{tab:hparams} in \hyperref[app:training_config]{Appendix~A} reports the final settings.

\textbf{Evaluation protocol.}
We evaluate each model on 10 runs for each of the 12 static benchmark scenarios (120 runs total). Each run starts in a fresh container and is capped at 60 rounds, following prior work on the same benchmark~\cite{happeKaplanCitoEMSE26}. The primary metric is the empirical success rate within 20 rounds, i.e., the sample estimate of $P(H_{\mathrm{root}}\le 20)$; this budget matches the \ac{RL} training horizon. Because every run records the round in which it first obtains root, success at other budgets, such as $r{=}5$ or $r{=}60$, is read from the same runs. We report Wilson 95\% confidence intervals throughout. All reported numbers are per-run probabilities under a fixed round budget, not best-of-$k$ retry metrics. The static benchmark is excluded from training and model selection (Section~\ref{sec:holdout} and Table~\ref{tab:training_selection_protocol}).

\textbf{Cost methodology.}
We estimate local inference cost for each local model from an empirical batched vLLM benchmark on an RTX~5090, and report API costs based on the recorded OpenRouter prices for the April--May 2026 evaluations~\cite{openrouter2026pricing}. We normalize all per-run cost figures to expected cost per successful root at $r{=}20$, computed as mean per-run cost, including failed attempts, divided by $P(H_{\mathrm{root}}\le 20)$. Training-stage compute (\ac{SFT} and \ac{RL}) is tracked separately as one-time post-training GPU-hours on H100 hardware, priced at European on-demand H100 rates~\cite{verda2026pricing}, and is not amortized into the per-run cost--reliability comparison; \hyperref[app:cost_methodology]{Appendix~A} gives the full equations.

\section{Results}
\label{sec:results}

Following the methodology described in Section~\ref{sec:methodology} and the experimental setup described in the previous section, we address the following research questions:

\begin{enumerate}
\item[\textbf{RQ1.}] Which supervised fine-tuning traces best warm-start a local security agent for \acl{RL}? [Section~\ref{sec:trace_ablation}]
\item[\textbf{RQ2.}] Which reward function design maximizes performance while avoiding reward-hacking behavior? [Section~\ref{sec:reward_ablation}]
\item[\textbf{RQ3.}] How competitive is the resulting local agent under a fixed budgeted protocol, and where does it fail? [Section~\ref{sec:main_results}]
\item[\textbf{RQ4.}] How cost-effective is the local agent against frontier APIs, and when does post-training amortize? [Section~\ref{sec:cost}]
\end{enumerate}

\subsection{Trace-Design Ablation}
\label{sec:trace_ablation}

We begin with the supervised stage that warm-starts the agent~(\textbf{RQ1}): which trace design produces the strongest initialization for downstream \ac{RL}?

\textbf{Setup.} We perform a $2{\times}3$ factorial ablation, varying two factors: trace-collection regime (guided vs.\ unguided) and reasoning format (none, short, or long). The ablation uses only procedural holdout performance for selection. Each condition trains the same \modelbackbone \ac{SFT} configuration on 2{,}000 balanced traces and evaluates 500 procedural holdout runs at the primary $r{\le}20$ budget.

\begin{table}[t]
  \centering
  \footnotesize
  \caption{Trace-design ablation on procedural holdout success at $r{\le}20$. Cells show success rate (successes/500); bold marks the selected configuration. \hyperref[tab:e1_trace_design]{Appendix Table~\ref*{tab:e1_trace_design}} gives Wilson intervals and trace statistics.}
  \label{tab:e1_trace_design_main}
  \setlength{\tabcolsep}{6pt}
  \begin{tabular}{lcc}
    \toprule
    Trace format & Guided & Unguided \\
    \midrule
    No reasoning    & 51.4\% (257/500) & 56.4\% (282/500) \\
    Short reasoning & 60.2\% (301/500) & 58.2\% (291/500) \\
    Long reasoning  & 65.2\% (326/500) & \textbf{73.2\% (366/500)} \\
    \bottomrule
  \end{tabular}
\end{table}

\textbf{Results.} Long reasoning is the best trace format in both regimes (Table~\ref{tab:e1_trace_design_main}), and unguided long reasoning is strongest overall at 73.2\%. It is at least 8.0 points above every other condition and 15.0 points above the unguided short-reasoning condition (\hyperref[tab:e1_trace_design]{Appendix Table~\ref*{tab:e1_trace_design}} gives Wilson 95\% intervals). Within the unguided regime, short reasoning adds little over no reasoning (58.2\% vs.\ 56.4\%), suggesting that retaining full observable-state reasoning matters more than adding a short reasoning field.

Giving the teacher hidden solution data does not produce the best student: the strongest traces come from a teacher that sees only the state available to the deployed student. Their open-ended exploration and recovery behavior appears to help the student more than solution-conditioned demonstrations. We therefore use unguided long-reasoning traces for \modelsft in all subsequent stages.\par\nobreak

\begin{findingbox}{Finding (RQ1). Use unguided teachers and full reasoning.}
Hidden solution data for the teacher does not give the best student. The best traces come from an unguided teacher that sees only what the student sees and keeps its full reasoning. The resulting student reaches 73.2\% on procedural holdout, at least 8 points above every other design.
\end{findingbox}

\subsection{Reward-Design Ablation}
\label{sec:reward_ablation}

With the warm-start fixed, we move to the \ac{RL} stage (\textbf{RQ2}): which reward improves reliability and efficiency without inducing reward hacking?

\textbf{Setup.} We train all four reward variants from the same \ac{SFT} checkpoint for 1{,}000 training steps on procedural environments and evaluate every 100 steps, using 100 holdout runs per checkpoint. Selection prioritizes success within 20 rounds, with success within 60 rounds and holdout success at step~1{,}000 as tie-breakers.

\textbf{Results.} Table~\ref{tab:reward_ablation} shows the selected checkpoint for each variant. The unshaped \rewardvariant{Outcome} reward already reaches 88.0\% success at $r{\le}20$, compared with 76.0\% for the \ac{SFT} initialization (228/300; Table~\ref{tab:e4_reward_ablation_details}). Both \rewardvariant{Outcome{+}Round} and \rewardvariant{Outcome{+}Cost} reach 90.0\%; cost shaping wins the extended-budget tie-breaker at 92.0\% versus 90.0\%. Round shaping is fastest early, reaching 80\% at $r{\le}5$ versus 67\% for cost shaping, but cost shaping catches up by $r{\le}20$.

At their peaks, the four rewards are close, between 87.0\% and 90.0\% (Table~\ref{tab:reward_ablation}). Holdout success then falls for every variant while training success stays high, which points to overfitting to the training scenarios~\cite{cobbeCoinRun}. The round bonus lowers success at step~1{,}000 by 8 points, whether added to the unshaped reward (83.0\% to 75.0\%) or to cost shaping (87.0\% to 79.0\%). A plausible reason is proxy optimization of the round count~\cite{skalseRewardHackingNeurIPS22}. The bonus rewards a shortcut~\cite{geirhosShortcutLearning2020} of trying familiar exploits before seeing evidence for them, as both round variants often do in their first round. Unseen hosts instead require acting on the evidence. We select \rewardvariant{Outcome{+}Cost} at step~300, as it wins both tie-breakers.

\begin{table}[t]
  \centering
  \caption{Reward-design ablation on procedural holdout (100 runs per checkpoint). Rows show the checkpoint selected by the procedural rule; $S_{\mathrm{all}}$ counts generator families solved on every run within 60 rounds. $^\star$Selected.}
  \label{tab:reward_ablation}
  \footnotesize\setlength{\tabcolsep}{5.5pt}
  \begin{tabular}{@{}lrrrrcr@{}}
    \toprule
    & & \multicolumn{3}{c}{Success (\%)} & & Step 1{,}000 \\
    \cmidrule(lr){3-5}
    Reward & Step & $r{\le}5$ & $r{\le}20$ & $r{\le}60$ & $S_{\mathrm{all}}$ & $r{\le}20$ (\%) \\
    \midrule
    \rewardvariant{Outcome}                & 200 & 50          & 88          & 90          & 6/10          & 83 \\
    \rewardvariant{{+}Round}               & 400 & \textbf{80} & \textbf{90} & 90          & 5/10          & 75 \\
    \rewardvariant{{+}Cost}$^\star$        & 300 & 67          & \textbf{90} & \textbf{92} & \textbf{7/10} & \textbf{87} \\
    \rewardvariant{{+}Round{+}Cost}        & 200 & 63          & 87          & 87          & 5/10          & 79 \\
    \bottomrule
  \end{tabular}
\end{table}

\begin{findingbox}{Finding (RQ2). Proxy efficiency rewards can teach shortcuts.}
For all rewards, training success stays high while holdout success falls. A round bonus gives the largest drop, from 90\% to 75\%, while a latency cost term ends at 87\%. With the round bonus, agents often try familiar exploits before seeing evidence, which transfers poorly.
\end{findingbox}

\subsection{Budgeted Static-Benchmark Results}
\label{sec:main_results}

\begin{figure}[t]
  \centering
  \includegraphics[width=\columnwidth]{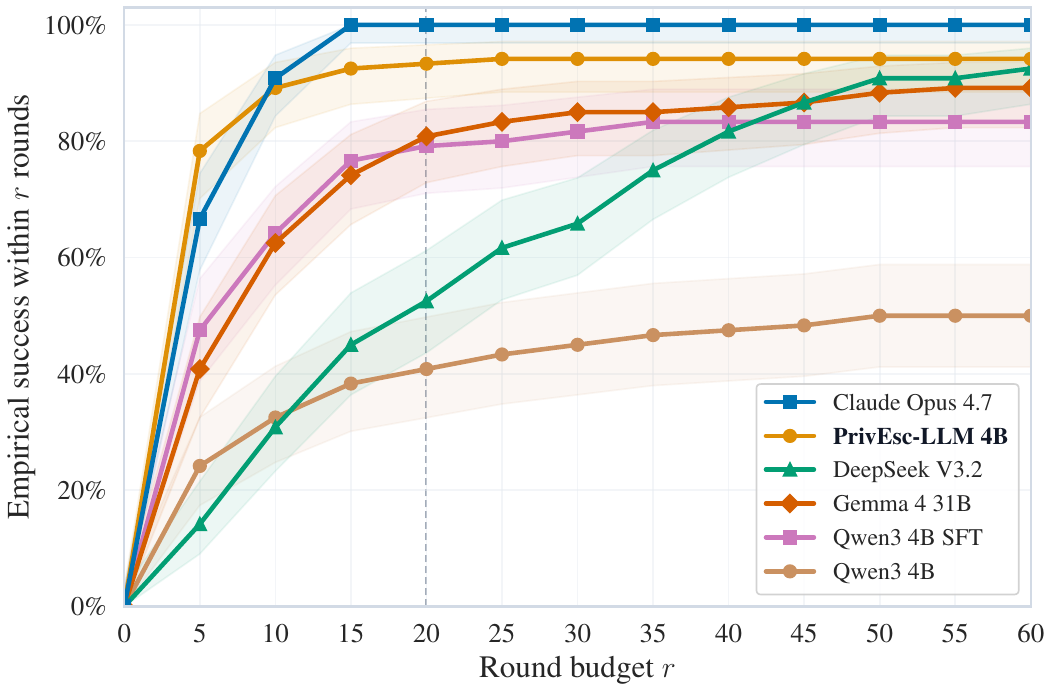}
  \caption{Empirical success within $r$ rounds across budgets. Each curve reports the fraction of runs that achieve root within $r$ rounds; shaded bands show 95\% Wilson CIs. Sample size is $N{=}120$ runs per model.}
  \label{fig:budget_curve}
\end{figure}

With trace and reward design fixed on procedural holdout, we turn to the static benchmark, the final external evaluation untouched by selection~(\textbf{RQ3}): how does the 4B policy compare to frontier API and larger open-weight baselines under tight budgets, and where does it fail?

\textbf{LLM results.} Figure~\ref{fig:budget_curve}
shows each model's success across round budgets on the static benchmark.
\modelbackbone reaches 40.8\% (49/120) success within 20 rounds. \ac{SFT} on the selected unguided
long-reasoning traces lifts \modelsft to 79.2\% (95/120), and \ac{RL} with
\rewardvariant{Outcome{+}Cost} further improves it to 93.3\% (112/120), yielding \modelrl.
While \modelclaude solves all 120 runs and sets the empirical ceiling, \modelrl outperforms it under a tight budget of $r{=}5$, reaching 78.3\% versus Claude's 66.7\%.
The two post-training stages contribute asymmetrically: \ac{SFT} provides the
large capability jump from the backbone, while \ac{RL} closes much, but not
all, of the remaining gap to Claude.

\textbf{Supporting baselines.}
The supporting evaluations in Table~\ref{tab:supplementary_baselines} use the same 120-run protocol and show that neither a larger model from the same family nor a similarly sized model from another family comes close to our post-trained models. The strongest additional local baseline, Qwen3 14B with thinking enabled, reaches 51.7\%, 27.5 points below \modelsft and 41.7 below \modelrl; Qwen3 8B reaches at most 30.8\% and the similarly sized Llama 3.2 3B only 9.2\%. The teacher, \modeldeepseekteacher, reaches 99.2\% at lower API cost than Claude, but its weights do not fit the consumer GPU used for local deployment (Section~\ref{sec:cost}).

\textbf{Non-LLM comparisons.}
Under a 30-minute per-scenario budget, \chainreactor returns a plan for 1 of 12 scenarios (\scenario{SUID GTFOBins}); three planner runs time out, and the remaining scenarios yield no plan. Its symbolic planner can only find chains that its fact extractor and planning domain can express, and several benchmark mechanisms, including credential discovery and container/host boundaries, fall outside that model.

Happe et~al.~\cite{happeKaplanCitoEMSE26} report approximately 75\% success for one professional penetration tester with seven years of experience, roughly five minutes per scenario, internet access, and arbitrary tools. They also report 8--25\% for traditional tools. These results on the same benchmark offer a useful reference point, although their time and tool budgets differ from our repeated-run protocol.

\begin{figure}[t]
  \centering
  \includegraphics[width=\columnwidth]{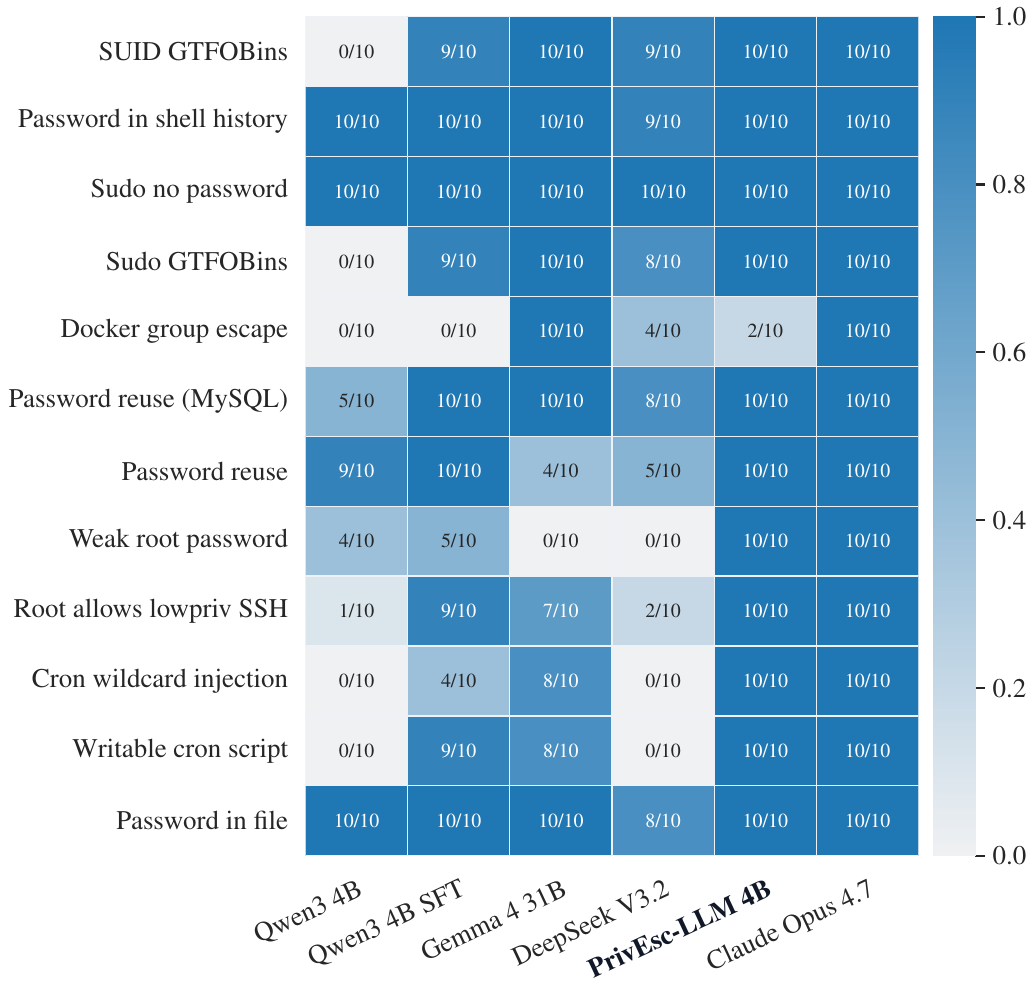}
  \caption{Per-scenario success at the primary $r{\le}20$ budget, where each cell shows $x/10$ successful runs and color encodes the success rate.}
  \label{fig:heatmap}
\end{figure}

\textbf{Per-scenario analysis.}
Figure~\ref{fig:heatmap} breaks down performance on a per-scenario basis. All counts use the same primary budget $r{\le}20$, where most models plateau.

Among our post-trained models, \modelsft already solves at least 9/10 runs on most scenarios but drops on \scenario{Weak root password} at 5/10, \scenario{Cron wildcard injection} at 4/10, and \scenario{Docker group escape} at 0/10. On the first two scenarios, it runs out of rounds because it tests too few password candidates or does not finish the multi-step cron exploit in time. \ac{RL} closes these gaps everywhere except Docker. \modelrl scores 10/10 on all 10 scenarios covered by procedural families, and 10/10 on \scenario{Sudo no password}, which the base model already solves. The sole remaining gap is \scenario{Docker group escape}, the withheld-family scenario, which we analyze below.

Among the external models, \modelclaude scores 10/10 across all 12 scenarios, including \scenario{Docker group escape}, the one scenario that \modelrl rarely solves. By contrast, \modeldeepseek achieves 10/10 on only 1 of 12 scenarios and fails entirely on three scenarios within 20~rounds, showing that frontier scale alone does not guarantee success under a tight round budget.

\textbf{Failure analysis.}
\modelrl fails 8 of 120 runs within 20 rounds, all on \scenario{Docker group escape}. Success on this scenario increases from 2/10 within 20 rounds to 3/10 within 60 rounds (\hyperref[app:examples]{Appendix~B} details one failed run).

The policy identifies Docker access in each failed run and usually obtains root inside a child container, but fails to turn this into verified root on the target host within 20 rounds. One run tries to mount the host filesystem with \mbox{\texttt{-v /:/host}}; none uses \texttt{chroot} through it. Repeated Docker launches and credential probes then consume the remaining rounds. The agent recognizes the mechanism but fails to complete the escape or recover from ineffective actions. A plausible explanation is limited Docker-specific knowledge in the backbone: this family provides no procedural training signal, and \ac{RL} tends to amplify behaviors already present after pretraining~\cite{zhaoEchoChamber}. The larger \modelgemma solves the same scenario 10/10 without our post-training, which suggests that success on this family depends strongly on pretrained knowledge.\par\nobreak

\begin{findingbox}{Finding (RQ3). Transfer holds within families, not across.}
On unseen instances of the trained families, \mbox{\modelrl} solves all 10 scenarios 10/10. For the withheld \mbox{\scenario{Docker group escape}} family it succeeds in only 2/10 runs, so transfer across families remains open. Overall, it reaches 93.3\% against 100\% for \mbox{\modelclaude}.
\end{findingbox}

\subsection{Training and Inference Cost}
\label{sec:cost}

With reliability settled, we finally turn to deployment economics~(\textbf{RQ4}): how does the local agent's cost--reliability operating point compare to API and larger local baselines, and when does post-training amortize?

\begin{figure}[t]
  \centering
  \includegraphics[width=\columnwidth]{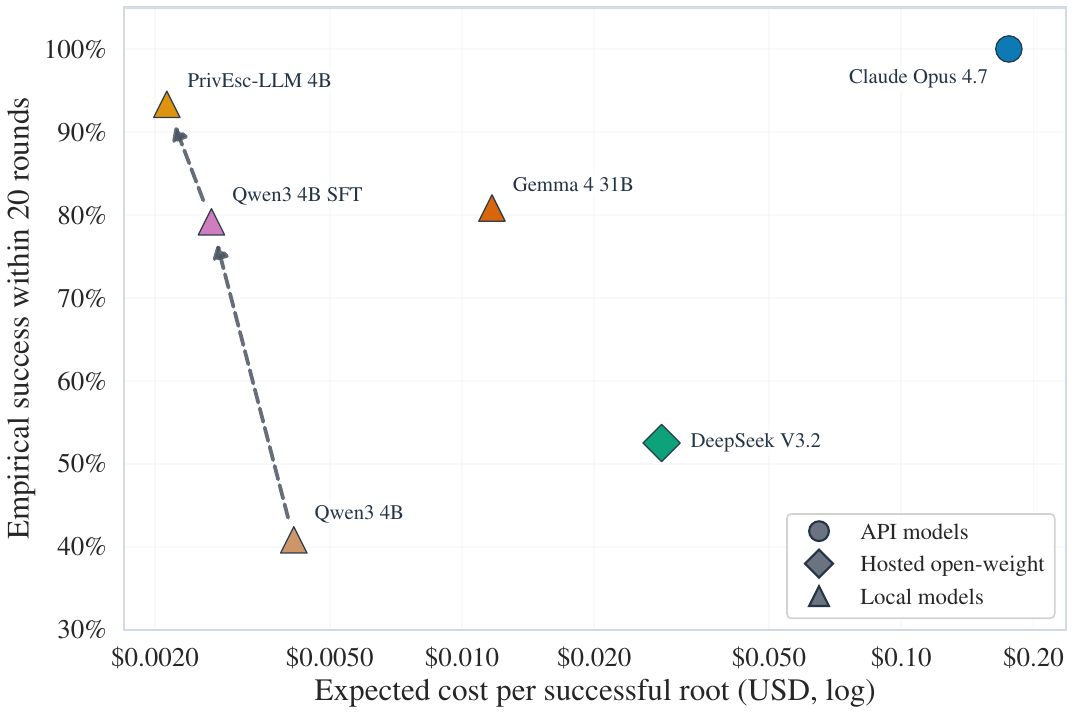}
  \caption{Static-benchmark reliability at $r{\le}20$ rounds versus expected cost per successful root at the same budget, over 120 runs per model. Cost is expected per-run cost divided by $P(H_{\mathrm{root}}\le 20)$, so the desirable region is upper left. Local points use effective input/output token prices anchored to measured batched serving cost; API points use provider token prices observed at evaluation time. The Qwen post-training progression moves from the backbone to \acs{SFT} and then \acs{RL}.}
  \label{fig:pareto_cost}
\end{figure}

\textbf{Results.} Post-training moves the Qwen3 4B models toward the upper-left region of Figure~\ref{fig:pareto_cost}: \modelbackbone reaches 40.8\% at \$0.00414 per successful root, \modelsft 79.2\% at \$0.00269, and \modelrl 93.3\% at \$0.00213. \modelrl trails \modelclaude by 6.7~points in reliability but is about 83$\times$ cheaper per successful root (\$0.00213 vs.\ \$0.1759), and 5.5$\times$ cheaper than \modelgemma while exceeding its 80.8\% success rate by 12.5~points. \modeldeepseek reaches 52.5\% at \$0.0285 per successful root, about 13$\times$ the cost of \modelrl. It recovers to 92.5\% by $r{=}60$ (Figure~\ref{fig:budget_curve}), but only with many more rounds than the primary budget allows.

Training the reported SFT model and RL checkpoint costs an estimated \$121.08, amortizing against \modelclaude after roughly 700 successful escalations at $r{=}20$. The teacher is the least expensive API alternative: \modeldeepseekteacher reaches 99.2\% at \$0.00469 per success, about 2.2$\times$ the cost of \modelrl. Its released weights~\cite{deepseekV4FlashModelCard}, however, occupy about 160~GB, roughly 20$\times$ the 8~GB of \modelrl and far beyond the 32~GB of the RTX~5090 consumer GPU used in our experiments. Among the tested models, \modelrl is the strongest that fits this GPU, and running locally keeps all host observations on the operator's machine.\par\nobreak

\begin{findingbox}{Finding (RQ4). Trained local agents are cost-effective.}
At 93.3\%, \mbox{\modelrl} is the cheapest tested model per success and reaches the upper-left region of the cost--reliability frontier. Compared with \mbox{\modelclaude}, it is about 83$\times$ cheaper, and its estimated \$121.08 training cost amortizes after roughly 700 successful escalations.
\end{findingbox}

\section{Discussion}
\label{sec:discussion}

Our results demonstrate that the proposed recipe can train strong local models for Linux privilege escalation. Below, we discuss what the ablations reveal, when local post-training pays off, how robust the trained agents are to prompting, and how the recipe extends to other tasks.

\textbf{Ablation insights.} Both ablations show which specific design choices within \ac{SFT} and \ac{RL} matter most. For \ac{SFT}, traces collected without hidden solution data but with full observable-state reasoning work best; in the unguided regime, short reasoning keeps only a small part of the gain that long reasoning has over no reasoning. For \ac{RL}, all rewards show signs of overfitting. Round shaping, alone or with cost shaping, ends below the unshaped \rewardvariant{Outcome} reward, whereas \rewardvariant{Outcome{+}Cost} ends highest.

\textbf{Cost amortization.} Post-training requires an upfront investment, whereas API-based frontier models are convenient to access. A local model, however, keeps data and deployment under the operator's control. Because the specialization is trained once and then reused, the investment amortizes quickly against frontier APIs. Under our local-accounting assumptions (on-demand H100 pricing~\cite{verda2026pricing}), the \$121.08 post-training cost is recovered after roughly 700 successful escalations against \modelclaude at $r{=}20$.

\textbf{Prompt sensitivity.} A post-hoc audit reruns the static benchmark with the detailed deployment prompt and two predeclared variants, using 240 new runs per model and prompt, without changing model selection or figures. The raw \modelbackbone is highly prompt-sensitive: at $r{\le}20$, success drops from 42.1\% with the detailed prompt to 25.4\% with the budget prompt and 21.7\% with the minimal prompt. The trained Qwen systems are much less sensitive in the same audit: \modelsft spans 75.8--80.8\% and \modelrl spans 88.3--91.7\%. The Qwen post-training gains therefore hold under all three prompts (Table~\ref{tab:prompt_sensitivity_audit}). \modelgemma is also stable (77.1--82.5\%), with the budget prompt slightly outperforming the detailed prompt.

\textbf{Beyond privilege escalation.} Although we focus on Linux privilege escalation, any task with a binary, environment-verifiable outcome and a tool-use loop is a plausible candidate for adoption, such as crash reproduction or patch validation~\cite{zhangAIxCCSoK26,wangCyberGymICLR26}. Linux privilege escalation is a convenient starting point: success is unambiguous, difficulty varies across exploit families, and the environment resets reliably, which lets us isolate the effect of each training design choice before moving to noisier domains. Because both post-training stages train only a lightweight \acs{LoRA} adapter on frozen backbone weights~\cite{lora}, each new domain can receive its own adapter instead of further fine-tuning a shared model. Earlier specializations are therefore not lost to catastrophic forgetting, and a single backbone copy can serve all adapters, so each further specialization adds little memory beyond the shared backbone~\cite{shengSLoRA}.

\section{Threats to Validity}
\label{sec:threats_to_validity}
\label{sec:threats}

Our evidence covers one 4B backbone specialized for Linux privilege escalation. Ten of the 12 benchmark scenarios share vulnerability families with training, and Docker is the only family withheld entirely, so transfer to other unseen families and security tasks remains to be tested. All benchmark hosts are exploitable: the root-owned canary rejects unsupported success claims, but false vulnerability reports, abstention, and stopping on hardened hosts are not measured. The historical human baseline uses different time and tool budgets, so a direct agent--human comparison would require a matched study. Finally, consumer hardware suffices for inference, whereas the \ac{RL} stage of the selected checkpoint required about 11.5 hours on four H100 GPUs.

\section{Related Work}
\label{sec:related}

\textbf{LLM agents for penetration testing.}
Prior \acs{LLM} pentest agents act in offensive-security loops but rarely combine local deployment, leakage-controlled generalization, and repeated-run budgeted evaluation. Early work showed that {\acs{LLM}}s can plan and execute autonomous Linux privilege escalation~\cite{happeCitoFSE23}, with follow-ups targeting enterprise Active Directory networks~\cite{happeCitoTOSEM25}. PentestGPT~\cite{dengPentestGPTUSENIX24} introduced a three-module agent and a real-world pentest benchmark. Closest to our setting is the controlled Linux privilege-escalation benchmark we build on~\cite{happeKaplanCitoEMSE26}, previously studied across cloud-hosted and local {\acs{LLM}}s. These agents prompt existing models. We instead post-train a small local model on leakage-controlled procedural environments and evaluate it with repeated runs under fixed budgets.

\textbf{Cybersecurity benchmarks.}
Most cybersecurity benchmarks emphasize breadth; fewer center value-disjoint generalization, fixed interaction budgets, repeated runs, and confidence intervals. CyBench~\cite{zhangCyBenchICLR25} evaluates agents on 40 \ac{CTF} tasks with subtask scoring, AutoPenBench~\cite{gioacchiniAutoPenBenchEMNLP25} introduces milestone-based scoring on 33 vulnerability-testing tasks, and CyberGym~\cite{wangCyberGymICLR26} scales execution-based reproduction to 1{,}507 real-world vulnerabilities across 188 projects. We focus on host-level Linux privilege escalation specifically because its success is automatically verifiable, which provides reliable reward signals and a budgeted primary metric.

\textbf{Post-training with verifiable rewards.}
\ac{RLVR} has improved multi-step reasoning on math and coding tasks when correctness can be verified automatically~\cite{deepseekmath,deepseekR1Nature,wenRLVRICLR26}. ToolRL~\cite{qianToolRLNeurIPS25} studies structured reward design for tool-use \ac{RL} and shows that coarse final-answer rewards are often insufficient for tool selection and parameterization. CTF-Dojo~\cite{zhuoCTFDojo} trains \ac{CTF} agents on execution-verified trajectories but relies on \ac{SFT} rather than \ac{RL}. Privilege escalation is a clean \ac{RLVR} setting because the trace shows whether the agent reached root, how many rounds it used, and whether its tool calls were valid. To the best of our knowledge, no prior work applies \ac{RLVR} to an interactive security task.

\section{Conclusion}
\label{sec:conclusion}

We presented a leakage-controlled post-training recipe that turns a small open-weight language model into a reliable tool-using security agent for Linux privilege escalation. Starting from \modelbackbone, supervised fine-tuning on unguided long-reasoning traces from procedurally generated environments raises success on the held-out benchmark from 40.8\% to 79.2\% within 20 rounds, and reinforcement learning with a cost-shaped verifiable reward lifts it to 93.3\%. These gains show that small local models can approach frontier-model reliability on a verifiable interactive security task. All remaining failures fall on the one vulnerability family withheld from training, which marks cross-family generalization as the main open challenge.

The resulting 4B agent is the strongest evaluated model that runs on a single consumer GPU. It trails \modelclaude by only 6.7 points at $r{=}20$, leads it at $r{=}5$ (78.3\% vs.\ 66.7\%), and costs about 83$\times$ less per successful root, so under our local-accounting assumptions the post-training cost pays for itself after roughly 700 successful escalations. Beyond privilege escalation, our recipe offers a practical path to reliable local agents for other security tasks with automatically verifiable outcomes, especially where sensitive host data must stay on the operator's own hardware.

\section{Ethical Considerations}
\label{sec:ethics}

This work trains a model to escalate privileges, so the dual-use risk is real. We mitigate that risk by restricting scenarios to already documented Linux misconfigurations and known privilege-escalation paths, running all experiments in isolated containers, and evaluating within a bounded benchmark rather than open-ended offensive tasks. The experiments do not interact with live third-party systems, collect personal data, or involve human subjects; accordingly, the work was outside the scope of institutional human-subjects review. Responsible disclosure to vendors is not applicable because the paper does not introduce new vulnerabilities or exploit primitives; its contribution is a post-training recipe for acting more reliably on public ones. Releasing the model and traces supports independent verification and defensive research on these documented techniques.

\section*{Acknowledgments}
This work was supported by the Vienna Science and Technology Fund (WWTF) under project BREADS (10.47379/VRG23011). Experiments ran on the MUSICA cluster provided by Austrian Scientific Computing (ASC).

\bibliographystyle{IEEEtran}
\bibliography{refs}

\begin{thebibliography}{10}
\providecommand{\url}[1]{#1}
\csname url@samestyle\endcsname
\providecommand{\newblock}{\relax}
\providecommand{\bibinfo}[2]{#2}
\providecommand{\BIBentrySTDinterwordspacing}{\spaceskip=0pt\relax}
\providecommand{\BIBentryALTinterwordstretchfactor}{4}
\providecommand{\BIBentryALTinterwordspacing}{\spaceskip=\fontdimen2\font plus
\BIBentryALTinterwordstretchfactor\fontdimen3\font minus
  \fontdimen4\font\relax}
\providecommand{\BIBforeignlanguage}[2]{{%
\expandafter\ifx\csname l@#1\endcsname\relax
\typeout{** WARNING: IEEEtran.bst: No hyphenation pattern has been}%
\typeout{** loaded for the language `#1'. Using the pattern for}%
\typeout{** the default language instead.}%
\else
\language=\csname l@#1\endcsname
\fi
#2}}
\providecommand{\BIBdecl}{\relax}
\BIBdecl

\bibitem{zhangAIxCCSoK26}
C.~Zhang, Y.~Park, F.~Fleischer, Y.-F. Fu, J.~Kim, D.~Kim, Y.~Kim, Q.~Xu,
  A.~Chin, Z.~Sheng \emph{et~al.}, ``{SoK}: {DARPA}'s {AI Cyber Challenge}
  ({AIxCC}): Competition design, architectures, and lessons learned,'' in
  \emph{Proc. of the {USENIX} Security Symposium}, 2026.

\bibitem{wangCyberGymICLR26}
Z.~Wang, T.~Shi, J.~He, M.~Cai, J.~Zhang, and D.~Song, ``{CyberGym}: Evaluating
  {AI} agents' real-world cybersecurity capabilities at scale,'' in \emph{Proc.
  of the Int. Conference on Learning Representations ({ICLR})}, 2026.

\bibitem{xiaFuzz4AllICSE24}
C.~S. Xia, M.~Paltenghi, J.~L. Tian, M.~Pradel, and L.~Zhang, ``{Fuzz4All}:
  Universal fuzzing with large language models,'' in \emph{Proc. of the
  {IEEE/ACM} Int. Conference on Software Engineering ({ICSE})}, 2024.

\bibitem{mozillaHardeningFirefox2026}
\BIBentryALTinterwordspacing
B.~Grinstead, C.~Holler, and F.~Braun, ``Behind the scenes hardening {Firefox}
  with {Claude Mythos Preview},'' Mozilla Hacks blog post, May 2026, accessed
  2026-05-25. [Online]. Available:
  \url{https://hacks.mozilla.org/2026/05/behind-the-scenes-hardening-firefox/}
\BIBentrySTDinterwordspacing

\bibitem{carliniExtractingTrainingData2021}
N.~Carlini, F.~Tram{\`e}r, E.~Wallace, M.~Jagielski, A.~Herbert-Voss, K.~Lee,
  A.~Roberts, T.~Brown, D.~Song, {\'U}.~Erlingsson, A.~Oprea, and C.~Raffel,
  ``Extracting training data from large language models,'' in \emph{Proc. of
  the {USENIX} Security Symposium}, 2021.

\bibitem{happeKaplanCitoEMSE26}
A.~Happe, A.~Kaplan, and J.~Cito, ``{LLMs} as hackers: Autonomous {Linux}
  privilege escalation attacks,'' \emph{Empirical Software Engineering},
  vol.~31, no.~3, 2026.

\bibitem{zhangCyBenchICLR25}
A.~K. Zhang, N.~Perry, R.~Dulepet, J.~Ji, C.~Menders, J.~W. Lin, E.~Jones,
  G.~Hussein, S.~Liu, D.~J. Jasper, P.~Peetathawatchai, A.~Glenn,
  V.~Sivashankar, D.~Zamoshchin, L.~Glikbarg, D.~Askaryar, H.~Yang, A.~Zhang,
  R.~Alluri, N.~Tran \emph{et~al.}, ``{CyBench}: A framework for evaluating
  cybersecurity capabilities and risks of language models,'' in \emph{Proc. of
  the Int. Conference on Learning Representations ({ICLR})}, 2025.

\bibitem{gtfobins}
\BIBentryALTinterwordspacing
{GTFOBins}, ``{GTFOBins},'' GitHub repository, 2026, accessed 2026-03-11.
  [Online]. Available: \url{https://github.com/GTFOBins/gtfobins.github.io}
\BIBentrySTDinterwordspacing

\bibitem{brownGPT3}
T.~B. Brown, B.~Mann, N.~Ryder, M.~Subbiah, J.~Kaplan, P.~Dhariwal,
  A.~Neelakantan, P.~Shyam, G.~Sastry, A.~Askell, S.~Agarwal, A.~Herbert-Voss,
  G.~Krueger, T.~Henighan, R.~Child, A.~Ramesh, D.~M. Ziegler, J.~Wu,
  C.~Winter, C.~Hesse, M.~Chen, E.~Sigler, M.~Litwin, S.~Gray, B.~Chess,
  J.~Clark, C.~Berner, S.~McCandlish, A.~Radford, I.~Sutskever, and D.~Amodei,
  ``Language models are few-shot learners,'' in \emph{Advances in Neural
  Information Processing Systems ({NeurIPS})}, 2020.

\bibitem{petroniLMKB}
F.~Petroni, T.~Rockt{\"a}schel, P.~Lewis, A.~Bakhtin, Y.~Wu, A.~H. Miller, and
  S.~Riedel, ``Language models as knowledge bases?'' in \emph{Proc. of the
  Conference on Empirical Methods in Natural Language Processing and the Int.
  Joint Conference on Natural Language Processing ({EMNLP-IJCNLP})}, 2019.

\bibitem{robertsParametricKnowledge}
A.~Roberts, C.~Raffel, and N.~Shazeer, ``How much knowledge can you pack into
  the parameters of a language model?'' in \emph{Proc. of the Conference on
  Empirical Methods in Natural Language Processing ({EMNLP})}, 2020.

\bibitem{changFactualKnowledge}
H.~Chang, J.~Park, S.~Ye, S.~Yang, Y.~Seo, D.-S. Chang, and M.~Seo, ``How do
  large language models acquire factual knowledge during pretraining?'' in
  \emph{Advances in Neural Information Processing Systems ({NeurIPS})}, 2024.

\bibitem{chungFlan}
H.~W. Chung, L.~Hou, S.~Longpre, B.~Zoph, Y.~Tay, W.~Fedus, Y.~Li, X.~Wang,
  M.~Dehghani, S.~Brahma, A.~Webson, S.~S. Gu, Z.~Dai, M.~Suzgun, X.~Chen,
  A.~Chowdhery, A.~Castro-Ros, M.~Pellat, K.~Robinson, D.~Valter, S.~Narang,
  G.~Mishra, A.~Yu, V.~Zhao, Y.~Huang, A.~Dai, H.~Yu, S.~Petrov, E.~H. Chi,
  J.~Dean, J.~Devlin, A.~Roberts, D.~Zhou, Q.~V. Le, and J.~Wei, ``Scaling
  instruction-finetuned language models,'' \emph{Journal of Machine Learning
  Research ({JMLR})}, vol.~25, 2024.

\bibitem{ouyangInstructGPT}
L.~Ouyang, J.~Wu, X.~Jiang, D.~Almeida, C.~Wainwright, P.~Mishkin, C.~Zhang,
  S.~Agarwal, K.~Slama, A.~Ray, J.~Schulman, J.~Hilton, F.~Kelton, L.~Miller,
  M.~Simens, A.~Askell, P.~Welinder, P.~F. Christiano, J.~Leike, and R.~Lowe,
  ``Training language models to follow instructions with human feedback,'' in
  \emph{Advances in Neural Information Processing Systems ({NeurIPS})}, 2022.

\bibitem{qwen3}
A.~Yang, A.~Li, B.~Yang \emph{et~al.}, ``{Qwen3} technical report,'' 2025,
  arXiv:2505.09388.

\bibitem{suttonBarto}
R.~S. Sutton and A.~G. Barto, \emph{Reinforcement Learning: An
  Introduction}.\hskip 1em plus 0.5em minus 0.4em\relax {MIT} Press, 2018.

\bibitem{christianoPreferences}
P.~F. Christiano, J.~Leike, T.~B. Brown, M.~Martic, S.~Legg, and D.~Amodei,
  ``Deep reinforcement learning from human preferences,'' in \emph{Advances in
  Neural Information Processing Systems ({NIPS})}, 2017.

\bibitem{deepseekmath}
Z.~Shao, P.~Wang, Q.~Zhu, R.~Xu, J.~Song, X.~Bi, H.~Zhang, M.~Zhang, Y.~K. Li,
  Y.~Wu, and D.~Guo, ``{DeepSeekMath}: Pushing the limits of mathematical
  reasoning in open language models,'' 2024, arXiv:2402.03300.

\bibitem{deepseekR1Nature}
{DeepSeek-AI}, D.~Guo, D.~Yang, H.~Zhang, J.~Song, R.~Zhang, R.~Xu, Q.~Zhu,
  S.~Ma, P.~Wang, X.~Bi \emph{et~al.}, ``{DeepSeek-R1} incentivizes reasoning
  in {LLM}s through reinforcement learning,'' \emph{Nature}, vol. 645, 2025.

\bibitem{wenRLVRICLR26}
X.~Wen, Z.~Liu, S.~Zheng, S.~Ye, Z.~Wu, Y.~Wang, Z.~Xu, X.~Liang, J.~Li,
  Z.~Miao, J.~Bian, and M.~Yang, ``Reinforcement learning with verifiable
  rewards implicitly incentivizes correct reasoning in base {LLM}s,'' in
  \emph{Proc. of the Int. Conference on Learning Representations ({ICLR})},
  2026.

\bibitem{rossDAgger}
S.~Ross, G.~Gordon, and D.~Bagnell, ``A reduction of imitation learning and
  structured prediction to no-regret online learning,'' in \emph{Proc. of the
  Int. Conference on Artificial Intelligence and Statistics ({AISTATS})}, 2011.

\bibitem{react}
S.~Yao, J.~Zhao, D.~Yu, N.~Du, I.~Shafran, K.~R. Narasimhan, and Y.~Cao,
  ``{ReAct}: Synergizing reasoning and acting in language models,'' in
  \emph{Proc. of the Int. Conference on Learning Representations ({ICLR})},
  2023.

\bibitem{toolformer}
T.~Schick, J.~Dwivedi-Yu, R.~Dess{\`i}, R.~Raileanu, M.~Lomeli, E.~Hambro,
  L.~Zettlemoyer, N.~Cancedda, and T.~Scialom, ``Toolformer: Language models
  can teach themselves to use tools,'' in \emph{Advances in Neural Information
  Processing Systems ({NeurIPS})}, 2023.

\bibitem{terminalBench}
M.~A. Merrill, A.~G. Shaw, N.~Carlini \emph{et~al.}, ``{Terminal-Bench}:
  Benchmarking agents on hard, realistic tasks in command line interfaces,'' in
  \emph{Proc. of the Int. Conference on Learning Representations ({ICLR})},
  2026.

\bibitem{cobbeProcgen}
K.~Cobbe, C.~Hesse, J.~Hilton, and J.~Schulman, ``Leveraging procedural
  generation to benchmark reinforcement learning,'' in \emph{Proc. of the Int.
  Conference on Machine Learning ({ICML})}, 2020.

\bibitem{cobbeCoinRun}
K.~Cobbe, O.~Klimov, C.~Hesse, T.~Kim, and J.~Schulman, ``Quantifying
  generalization in reinforcement learning,'' in \emph{Proc. of the Int.
  Conference on Machine Learning ({ICML})}, 2019.

\bibitem{arpDosDonts}
D.~Arp, E.~Quiring, F.~Pendlebury, A.~Warnecke, F.~Pierazzi, C.~Wressnegger,
  L.~Cavallaro, and K.~Rieck, ``Dos and don{\textquoteright}ts of machine
  learning in computer security,'' in \emph{Proc. of the {USENIX} Security
  Symposium}, 2022.

\bibitem{evertzChasingShadows}
J.~Evertz, N.~Risse, N.~Neuer, A.~M{\"u}ller, P.~Normann, G.~Sapia, S.~Gupta,
  D.~Pape, S.~Shaw, D.~Srivastav, C.~Wressnegger, E.~Quiring, T.~Eisenhofer,
  D.~Arp, and L.~Sch{\"o}nherr, ``Chasing shadows: Pitfalls in {LLM} security
  research,'' in \emph{Proc. of the Network and Distributed System Security
  Symposium ({NDSS})}, 2026.

\bibitem{dengDataContamination2024}
C.~Deng, Y.~Zhao, X.~Tang, M.~Gerstein, and A.~Cohan, ``Investigating data
  contamination in modern benchmarks for large language models,'' in
  \emph{Proc. of the Conference of the North American Chapter of the
  Association for Computational Linguistics: Human Language Technologies
  ({NAACL-HLT})}, 2024.

\bibitem{deepseekV4FlashModelCard}
\BIBentryALTinterwordspacing
{DeepSeek-AI}, ``{DeepSeek-V4-Flash} model card,'' Hugging Face model card,
  2026, accessed 2026-09-30. [Online]. Available:
  \url{https://huggingface.co/deepseek-ai/DeepSeek-V4-Flash}
\BIBentrySTDinterwordspacing

\bibitem{lora}
E.~J. Hu, Y.~Shen, P.~Wallis, Z.~Allen-Zhu, Y.~Li, S.~Wang, L.~Wang, and
  W.~Chen, ``{LoRA}: Low-rank adaptation of large language models,'' in
  \emph{Proc. of the Int. Conference on Learning Representations ({ICLR})},
  2022.

\bibitem{unsloth}
\BIBentryALTinterwordspacing
D.~Han, M.~Han, and {Unsloth team}, ``{Unsloth},'' GitHub repository, 2023,
  accessed 2026-03-05. [Online]. Available:
  \url{https://github.com/unslothai/unsloth}
\BIBentrySTDinterwordspacing

\bibitem{lorawr}
J.~Schulman and {Thinking Machines Lab}, ``{LoRA} without regret,''
  \emph{Thinking Machines Lab: Connectionism}, 2025.

\bibitem{gudibandeFalsePromise2024}
A.~Gudibande, E.~Wallace, C.~V. Snell, X.~Geng, H.~Liu, P.~Abbeel, S.~Levine,
  and D.~Song, ``The false promise of imitating proprietary language models,''
  in \emph{Proc. of the Int. Conference on Learning Representations ({ICLR})},
  2024.

\bibitem{ngRewardShaping}
A.~Y. Ng, D.~Harada, and S.~J. Russell, ``Policy invariance under reward
  transformations: Theory and application to reward shaping,'' in \emph{Proc.
  of the Int. Conference on Machine Learning ({ICML})}, 1999.

\bibitem{chenDEPOAAAI26}
S.~Chen, M.~Zhao, L.~Xu, Y.~Zhao, B.~Zhu, H.~Zhang, S.~Zhao, and C.~Lu,
  ``{DEPO}: Dual-efficiency preference optimization for {LLM} agents,'' in
  \emph{Proc. of the {AAAI} Conference on Artificial Intelligence ({AAAI})},
  2026.

\bibitem{skalseRewardHackingNeurIPS22}
J.~Skalse, N.~H.~R. Howe, D.~Krasheninnikov, and D.~Krueger, ``Defining and
  characterizing reward gaming,'' in \emph{Advances in Neural Information
  Processing Systems ({NeurIPS})}, 2022.

\bibitem{qianToolRLNeurIPS25}
C.~Qian, E.~C. Acikgoz, Q.~He, H.~Wang, X.~Chen, D.~Hakkani-T{\"u}r, G.~Tur,
  and H.~Ji, ``{ToolRL}: Reward is all tool learning needs,'' in \emph{Advances
  in Neural Information Processing Systems ({NeurIPS})}, 2025.

\bibitem{primerl}
\BIBentryALTinterwordspacing
{Prime Intellect}, ``{PRIME-RL},'' GitHub repository, 2025, accessed
  2026-03-04. [Online]. Available:
  \url{https://github.com/PrimeIntellect-ai/prime-rl}
\BIBentrySTDinterwordspacing

\bibitem{verifiers}
\BIBentryALTinterwordspacing
W.~Brown, ``{Verifiers}: Environments for {LLM} reinforcement learning,''
  GitHub repository, 2025, accessed 2026-03-05. [Online]. Available:
  \url{https://github.com/PrimeIntellect-ai/verifiers}
\BIBentrySTDinterwordspacing

\bibitem{kwonPagedAttention}
W.~Kwon, Z.~Li, S.~Zhuang, Y.~Sheng, L.~Zheng, C.~H. Yu, J.~E. Gonzalez,
  H.~Zhang, and I.~Stoica, ``Efficient memory management for large language
  model serving with {PagedAttention},'' in \emph{Proc. of the {ACM} Symposium
  on Operating Systems Principles ({SOSP})}, 2023.

\bibitem{gemma4ModelCard}
\BIBentryALTinterwordspacing
{Google DeepMind}, ``{Gemma 4 31B IT} model card,'' Hugging Face model card,
  2026, accessed 2026-05-22. [Online]. Available:
  \url{https://huggingface.co/google/gemma-4-31B-it}
\BIBentrySTDinterwordspacing

\bibitem{deepseekV3_2}
{DeepSeek-AI}, A.~Liu, A.~Mei \emph{et~al.}, ``{DeepSeek-V3.2}: Pushing the
  frontier of open large language models,'' 2025, arXiv:2512.02556.

\bibitem{claudeOpus47SystemCard}
\BIBentryALTinterwordspacing
{Anthropic}, ``{Claude Opus 4.7} system card,'' Anthropic system card, 2026,
  accessed 2026-09-30. [Online]. Available:
  \url{https://www.anthropic.com/claude-opus-4-7-system-card}
\BIBentrySTDinterwordspacing

\bibitem{openrouter2026pricing}
\BIBentryALTinterwordspacing
{OpenRouter}, ``Pricing,'' Pricing page, 2026, accessed 2026-05-04. [Online].
  Available: \url{https://openrouter.ai/pricing}
\BIBentrySTDinterwordspacing

\bibitem{llama32ModelCard}
\BIBentryALTinterwordspacing
{Meta}, ``{Llama 3.2 3B Instruct} model card,'' Hugging Face model card, 2024,
  accessed 2026-09-30. [Online]. Available:
  \url{https://huggingface.co/meta-llama/Llama-3.2-3B-Instruct}
\BIBentrySTDinterwordspacing

\bibitem{dePasqualeChainReactorUSENIX24}
G.~{De Pasquale}, I.~Grishchenko, R.~Iesari, G.~Pizarro, L.~Cavallaro,
  C.~Kruegel, and G.~Vigna, ``{ChainReactor}: Automated privilege escalation
  chain discovery via {AI} planning,'' in \emph{Proc. of the {USENIX} Security
  Symposium}, 2024.

\bibitem{traitor}
\BIBentryALTinterwordspacing
L.~Galvin, ``Traitor: Automatic {Linux} privilege escalation,'' GitHub
  repository, 2021, accessed 2026-03-10. [Online]. Available:
  \url{https://github.com/liamg/traitor}
\BIBentrySTDinterwordspacing

\bibitem{pwncatcs}
\BIBentryALTinterwordspacing
C.~Stewart, ``{pwncat},'' GitHub repository, 2021, accessed 2026-03-10.
  [Online]. Available: \url{https://github.com/calebstewart/pwncat}
\BIBentrySTDinterwordspacing

\bibitem{qwen3FourBInstruct2507ModelCard}
\BIBentryALTinterwordspacing
{Qwen Team}, ``{Qwen3-4B-Instruct-2507} model card,'' Hugging Face model card,
  2025, accessed 2026-05-25. [Online]. Available:
  \url{https://huggingface.co/Qwen/Qwen3-4B-Instruct-2507}
\BIBentrySTDinterwordspacing

\bibitem{deepseekV3_2ModelCard}
\BIBentryALTinterwordspacing
{DeepSeek-AI}, ``{DeepSeek-V3.2} model card,'' Hugging Face model card, 2025,
  accessed 2026-05-25. [Online]. Available:
  \url{https://huggingface.co/deepseek-ai/DeepSeek-V3.2}
\BIBentrySTDinterwordspacing

\bibitem{dettmersQLoRA}
T.~Dettmers, A.~Pagnoni, A.~Holtzman, and L.~Zettlemoyer, ``{QLoRA}: Efficient
  finetuning of quantized {LLM}s,'' in \emph{Advances in Neural Information
  Processing Systems ({NeurIPS})}, 2023.

\bibitem{verda2026pricing}
\BIBentryALTinterwordspacing
{Verda}, ``Premium {GPU} instances,'' Pricing page, 2026, accessed 2026-03-07.
  [Online]. Available:
  \url{https://web.archive.org/web/20260304072228/https://verda.com/products}
\BIBentrySTDinterwordspacing

\bibitem{geirhosShortcutLearning2020}
R.~Geirhos, J.-H. Jacobsen, C.~Michaelis, R.~Zemel, W.~Brendel, M.~Bethge, and
  F.~A. Wichmann, ``Shortcut learning in deep neural networks,'' \emph{Nature
  Machine Intelligence}, vol.~2, no.~11, 2020.

\bibitem{zhaoEchoChamber}
R.~Zhao, A.~Meterez, S.~M. Kakade, C.~Pehlevan, S.~Jelassi, and E.~Malach,
  ``Echo chamber: {RL} post-training amplifies behaviors learned in
  pretraining,'' in \emph{Proc. of the Conference on Language Modeling
  ({COLM})}, 2025.

\bibitem{shengSLoRA}
Y.~Sheng, S.~Cao, D.~Li \emph{et~al.}, ``{S-LoRA}: Serving thousands of
  concurrent {LoRA} adapters,'' in \emph{Proc. of the Conference on Machine
  Learning and Systems ({MLSys})}, 2024.

\bibitem{happeCitoFSE23}
A.~Happe and J.~Cito, ``Getting pwn'd by {AI}: Penetration testing with large
  language models,'' in \emph{Proc. of the {ACM} Joint European Software
  Engineering Conference and Symposium on the Foundations of Software
  Engineering ({ESEC/FSE})}, 2023.

\bibitem{happeCitoTOSEM25}
------, ``Can {LLM}s hack enterprise networks? {A}utonomous assumed breach
  penetration-testing {Active Directory} networks,'' \emph{{ACM} Transactions
  on Software Engineering and Methodology}, vol.~35, no.~6, 2026.

\bibitem{dengPentestGPTUSENIX24}
G.~Deng, Y.~Liu, V.~Mayoral-Vilches, P.~Liu, Y.~Li, Y.~Xu, T.~Zhang, Y.~Liu,
  M.~Pinzger, and S.~Rass, ``{PentestGPT}: Evaluating and harnessing large
  language models for automated penetration testing,'' in \emph{Proc. of the
  {USENIX} Security Symposium}, 2024.

\bibitem{gioacchiniAutoPenBenchEMNLP25}
L.~Gioacchini, A.~Delsanto, I.~Drago, M.~Mellia, G.~Siracusano, and R.~Bifulco,
  ``{AutoPenBench}: A vulnerability testing benchmark for generative agents,''
  in \emph{Proc. of the Conference on Empirical Methods in Natural Language
  Processing: Industry Track ({EMNLP Industry Track})}, 2025.

\bibitem{zhuoCTFDojo}
T.~Y. Zhuo, D.~Wang, H.~Ding, V.~Kumar, and Z.~Wang, ``Training language model
  agents to find vulnerabilities with {CTF-Dojo},'' in \emph{Proc. of the Int.
  Conference on Machine Learning ({ICML})}, 2026.

\bibitem{eurostatElectricityH22025}
\BIBentryALTinterwordspacing
{Eurostat}, ``Electricity prices for non-household consumers, second half
  2025,'' Dataset \texttt{nrg\_pc\_205}, EU average including non-recoverable
  taxes and levies, 2026, accessed 2026-05-24. [Online]. Available:
  \url{https://ec.europa.eu/eurostat/databrowser/view/nrg_pc_205/default/table}
\BIBentrySTDinterwordspacing

\end{thebibliography}

\appendices

\newcommand{\appendixsubheading}[2]{%
  \par\addvspace{0.65\baselineskip}%
  \needspace{4\baselineskip}%
  \phantomsection
  \noindent\textbf{#1}\label{#2}\par\nobreak\vspace{0.15\baselineskip}\noindent\ignorespaces
}

\section{Additional Reproducibility Details}
\label{app:repro_details}
This appendix gives the generators and benchmark exclusions, the reward, training, and inference settings, the supporting model comparisons, and the cost methodology.

\begin{table}[!htbp]
  \centering
  \caption{Procedural generator families used in training and validation. The table summarizes the 10 seeded procedural environments used for trace collection, \acs{SFT} data assembly, and \acs{RL} training. Per-scenario benchmark exclusions are listed in Table~\ref{tab:scenario_taxonomy}.}
  \label{tab:generator_catalog}
  \scriptsize
  \begin{tabular}{@{}>{\raggedright\arraybackslash}p{0.30\columnwidth}>{\raggedright\arraybackslash}p{0.65\columnwidth}@{}}
    \toprule
    Generator & Description \\
    \midrule
    SUID GTFOBins & Places a non-interactive GTFOBins binary on disk with the SUID bit set. \\
    Sudo GTFOBins & Grants passwordless sudo access to a non-interactive GTFOBins binary. \\
    File capabilities & Assigns Linux file capabilities to a GTFOBins binary. \\
    Password-history leakage & Leaks reusable credentials through shell history. \\
    Password-file leakage & Stores reusable credentials in files. \\
    Password reuse & Reuses a low-privilege credential as the root password. \\
    Weak root password & Samples the root password from a weak-password distribution. \\
    Cron wildcard injection & Creates a cron-driven archive workflow vulnerable to wildcard injection. \\
    Writable cron script & Exposes a writable cron-executed script. \\
    SSH key reuse & Reuses an existing private key for root authentication. \\
    \bottomrule
  \end{tabular}
\end{table}

\begin{table*}[t]
  \centering
  \caption{Benchmark scenario taxonomy: the 12 static benchmark scenarios, mapped to their closest procedural generator family (Table~\ref{tab:generator_catalog}) and labeled \textnormal{\textit{covered}} (vulnerability class in training, benchmark-specific values excluded), \textnormal{\textit{no direct procedural analog}}, or \textnormal{\textit{withheld family}}. All scenarios also inherit the universal \textnormal{\texttt{lowpriv}} and \textnormal{\texttt{/home/lowpriv/}} exclusions.}
  \label{tab:scenario_taxonomy}
  \scriptsize
  \begin{tabular}{@{}>{\raggedright\arraybackslash}p{0.18\textwidth}>{\raggedright\arraybackslash}p{0.18\textwidth}>{\raggedright\arraybackslash}p{0.16\textwidth}>{\raggedright\arraybackslash}p{0.40\textwidth}@{}}
    \toprule
    Static benchmark scenario & Closest procedural family & Distribution status & Benchmark-specific values excluded from sampling or audited absent \\
    \midrule
    SUID GTFOBins              & SUID GTFOBins              & covered                       & Binaries \texttt{find}, \texttt{python}, \texttt{python3}, \texttt{python3.11} \\
    Password in shell history  & Password-history leakage   & covered                       & Password \texttt{aim8Du7h}; pair \texttt{root:aim8Du7h}; three exact history-leakage snippets (\texttt{sshpass}, \texttt{echo | su root}, \texttt{expect}) \\
    Sudo no password           & ---                        & no direct procedural analog   & No generator samples unrestricted \texttt{NOPASSWD:ALL}, as the restricted \scenario{Sudo GTFOBins} family is separate; sudoers filename \texttt{allow-all}; exact \texttt{NOPASSWD:ALL} string \\
    Sudo GTFOBins              & Sudo GTFOBins              & covered                       & Binary \texttt{tar}; sudoers filename \texttt{/etc/sudoers.d/gtfo-tar}; \texttt{NOPASSWD:/usr/bin/tar} string \\
    Docker group escape        & ---                        & withheld family               & Whole family withheld, so no instance exposes docker-group membership or a Docker socket; host-bind-mount solve fragments \\
    Password reuse (MySQL)     & Password-file leakage      & covered                       & Password \texttt{aim8Du7h}; pair \texttt{root:aim8Du7h}; filename \texttt{.my.cnf} \\
    Password reuse             & Password reuse             & covered                       & Password \texttt{trustno1}; pair \texttt{root:trustno1} \\
    Weak root password         & Weak root password         & covered                       & Password \texttt{root} removed from the common-password pool \\
    Root allows lowpriv SSH    & SSH key reuse              & covered                       & SSH key name \texttt{id\_rsa} \\
    Cron wildcard injection    & Cron wildcard injection    & covered                       & Cron job name \texttt{backup\_lowpriv}; \texttt{backup} removed from \texttt{cron\_backup\_dirs}; \texttt{/home/*/backup} paths \\
    Writable cron script       & Writable cron script       & covered                       & Cron job name \texttt{lowpriv\_script}; \texttt{cron.sh} removed from \texttt{cron\_script\_names} \\
    Password in file           & Password-file leakage      & covered                       & Password \texttt{aim8Du7h}; pair \texttt{root:aim8Du7h}; filename \texttt{admin\_on\_vacation.txt} \\
    \bottomrule
  \end{tabular}
\end{table*}

\appendixsubheading{Reward Function Details}{app:reward_details}
The four reward variants are defined in Equation~\ref{eq:reward_variants}. We select \rewardvariant{Outcome{+}Cost} for \modelrl based on procedural holdout (Section~\ref{sec:reward_ablation}). All variants use $h_{\max}{=}20$ and $\lambda_{\mathrm{iface}}{=}0.05$; cost-shaped variants use $\lambda_{\mathrm{cost}}{=}0.1$. For the cost term $C(e)$ in Equation~\ref{eq:cost_term}, we use latency clips of $c^{\text{LLM}}_{\mathrm{clip}}{=}20$~s and $c^{\text{tool}}_{\mathrm{clip}}{=}65$~s, with $C_{\mathrm{ref}}{=}540$~s, calibrated as described below.

\appendixsubheading{Reward-Cost Calibration}{app:reward_cost_calibration}
The latency clips and reference cost are calibrated before \ac{RL} using a timing benchmark of the selected \ac{SFT} adapter under training-like H100 concurrency (100 traces, 74 successful). Success-conditioned per-call latencies have \acs{LLM} $P_{99}{=}18.1$~s and cron-scenario tool $P_{99}{=}30.9$~s, motivating $c^{\text{LLM}}_{\mathrm{clip}}{=}20$~s and $c^{\text{tool}}_{\mathrm{clip}}{=}65$~s; these clip 0.60\% of successful \acs{LLM} calls and 0.11\% of tool calls. Successful clipped episode cost has $P_{99}{=}272$~s, so $C_{\mathrm{ref}}{=}540$~s (${\approx}2{\times}P_{99}$, rounded to 10~s) allows additional tool calls before the cost bonus reaches zero; 0/74 calibration successes saturate, and post-hoc training rollouts show a 0.38\% saturation rate.

\phantomsection
\label{app:training_config}
\begin{table}[!htbp]
  \centering
  \caption{Training-stage selection protocol used before static-benchmark evaluation.}
  \label{tab:training_selection_protocol}
  \footnotesize\setlength{\tabcolsep}{3pt}
  \begin{tabular}{@{}lp{0.72\columnwidth}@{}}
    \toprule
    Stage & Selection protocol \\
    \midrule
    \acs{SFT} &
    \raggedright Validation-loss filter, then procedural holdout success. Phase~A LR sweep at rank~64: $1.5{\times}10^{-4}$, $5{\times}10^{-4}$, $1.5{\times}10^{-3}$; Phase~B rank sweep: 4, 8, 16, 32, 64. Selected $\text{LR}{=}1.5{\times}10^{-4}$, rank~8; confirmed across three seeds (76.1\%\,$\pm$\,0.8\,pp at epoch~10). \tabularnewline
    \acs{RL} &
    \raggedright Root success on procedural training rollouts at step~200. LR sweep: $3.75{\times}10^{-6}$, $7.5{\times}10^{-6}$, $1.5{\times}10^{-5}$, $3{\times}10^{-5}$, $6{\times}10^{-5}$. Selected $\text{LR}{=}1.5{\times}10^{-5}$ from the selected \ac{SFT} checkpoint, keeping rank and $\alpha$. \tabularnewline
    \bottomrule
  \end{tabular}
\end{table}

\phantomsection
\label{app:inference_config}
\begin{table}[!htbp]
  \centering
  \caption{Inference settings for the primary comparison models. API models are accessed via OpenRouter~\cite{openrouter2026pricing}.}
  \label{tab:inference_config}
  \scriptsize\setlength{\tabcolsep}{2pt}
  \begin{tabular}{lcccc}
    \toprule
    & Qwen3 4B & Gemma & DeepSeek V3.2 & Claude \\
    & local & local & API & API \\
    \midrule
    Weight format  & bf16\textsuperscript{\S} & 4-bit BnB & --- & --- \\
    Temperature    & 0.7      & 0.7      & default\textsuperscript{\textdagger} & default\textsuperscript{\textdaggerdbl} \\
    top-p          & 0.8      & 0.95     & default\textsuperscript{\textdagger} & default\textsuperscript{\textdaggerdbl} \\
    top-k          & 20       & ---      & ---     & ---     \\
    Max output     & 2{,}048  & 2{,}048  & 2{,}048 & 2{,}048 \\
    Context cap    & 32K      & 32K\textsuperscript{\P} & 128K    & 200K    \\
    Reasoning      & ---      & thinking & high    & high    \\
    Parser         & Hermes   & Gemma4   & ---     & ---     \\
    \bottomrule
    \multicolumn{5}{@{}p{0.96\columnwidth}@{}}{\scriptsize\textsuperscript{\S}Qwen \acs{SFT} used 4-bit \acs{QLoRA}; static inference used bf16 vLLM weights.} \\
    \multicolumn{5}{@{}p{0.96\columnwidth}@{}}{\scriptsize\textsuperscript{\P}Gemma local serving used \nolinkurl{unsloth/gemma-4-31B-it-unsloth-bnb-4bit}; model-card maximum is 256K, capped to 32K to fit an RTX~5090-class memory budget.} \\
    \multicolumn{5}{@{}p{0.96\columnwidth}@{}}{\scriptsize\textsuperscript{\textdagger}$T{=}1.0$, top-p${=}0.95$. \textsuperscript{\textdaggerdbl}$T{=}1.0$, top-p${=}1.0$.}
  \end{tabular}
\end{table}

\begin{table}[!htbp]
  \centering
  \caption{Hyperparameters for both training stages.}
  \label{tab:hparams}
  \footnotesize\setlength{\tabcolsep}{3pt}
  \begin{tabular}{@{}lll@{}}
    \toprule
    & \acs{SFT} & \acs{RL} \\
    \midrule
    Learning rate & $1.5{\times}10^{-4}$ & $1.5{\times}10^{-5}$ \\
    LR scheduler & linear (no warmup) & constant \\
    Optimizer & \acs{AdamW} 8-bit ($\beta_2{=}0.95$) & \acs{AdamW} \\
    \acs{LoRA} rank / alpha & 8\,/\,32 & 8\,/\,32 \\
    \acs{LoRA} targets & \multicolumn{2}{c}{all linear + \texttt{lm\_head}$^\dagger$} \\
    \acs{LoRA} dropout & 0.0 & 0.0 \\
    Precision & \acs{QLoRA} (4-bit) & bf16 \\
    Sequence length & 32{,}768 & 32{,}768 \\
    Batch size & 8 & 80 (rollouts: 8) \\
    Checkpoint / steps & epoch~10 of~10 & step 300 of 1{,}000 \\
    Training horizon (rounds) & --- & 20 \\
    Hardware & 4$\times$H100, ${\approx}1$\,hr~$40$\,min & 4$\times$H100, ${\approx}11.5$\,hr \\
    \bottomrule
    \multicolumn{3}{@{}l@{}}{\scriptsize $^\dagger$\texttt{q,k,v,o\_proj}, \texttt{gate,up,down\_proj}, \texttt{lm\_head}}
  \end{tabular}
\end{table}

\begin{table*}[t]
  \centering
  \caption{Trace-design ablation on procedural holdout environments. Each row trains \textnormal{\modelbackbone} with the same \acs{SFT} hyperparameters on 2{,}000 traces and evaluates 500 holdout runs. Success cells show percent (successes/500); Wilson intervals are run-level and exclude training-seed variation. The static benchmark is not used for this choice.}
  \label{tab:e1_trace_design}
  \footnotesize\setlength{\tabcolsep}{3pt}
  \begin{tabular}{llrrrcr}
    \toprule
    Collection & Trace format & Avg. tokens/trace & Traces $>$16K & Success $r{\le}20$ & 95\% CI & Success $r{\le}60$ \\
    \midrule
    Guided & No reasoning & 3{,}122 & 0 & 51.4\% (257/500) & 47.0--55.8 & 52.0\% (260/500) \\
    Guided & Short reasoning & 3{,}213 & 0 & 60.2\% (301/500) & 55.8--64.4 & 60.4\% (302/500) \\
    Guided & Long reasoning & 3{,}525 & 0 & 65.2\% (326/500) & 60.9--69.2 & 66.4\% (332/500) \\
    Unguided & No reasoning & 7{,}314 & 119 & 56.4\% (282/500) & 52.0--60.7 & 61.0\% (305/500) \\
    Unguided & Short reasoning & 7{,}488 & 126 & 58.2\% (291/500) & 53.8--62.4 & 58.4\% (292/500) \\
    Unguided & Long reasoning & 8{,}751 & 196 & \textbf{73.2\% (366/500)} & \textbf{69.2--76.9} & \textbf{74.8\% (374/500)} \\
    \bottomrule
  \end{tabular}
\end{table*}

\appendixsubheading{Supporting Model Comparisons}{app:supplementary_baselines}
Table~\ref{tab:supplementary_baselines} reports the additional static evaluations. All use the deployment prompt, the same tools and root verifier, 10 runs per scenario, and a 60-round cap; the table reports success within 20 rounds.

\begin{table}[!htbp]
  \centering
  \caption{Supporting static-benchmark evaluations. Each row contains 120 runs; intervals are Wilson 95\% CIs.}
  \label{tab:supplementary_baselines}
  \scriptsize\setlength{\tabcolsep}{3pt}
  \begin{tabular}{@{}llrr@{}}
    \toprule
    Model & Reasoning & Success & 95\% CI \\
    \midrule
    Qwen3 8B & on & 30.8\% (37/120) & 23.3--39.6 \\
    Qwen3 8B & off & 20.0\% (24/120) & 13.8--28.0 \\
    Qwen3 14B (FP8) & on & 51.7\% (62/120) & 42.8--60.4 \\
    Qwen3 14B (FP8) & off & 27.5\% (33/120) & 20.3--36.1 \\
    Llama 3.2 3B Instruct & n/a & 9.2\% (11/120) & 5.2--15.7 \\
    DeepSeek V4 Flash & high & 99.2\% (119/120) & 95.4--99.9 \\
    \bottomrule
  \end{tabular}
\end{table}

The local checkpoints are \path{Qwen/Qwen3-8B}, \path{Qwen/Qwen3-14B-FP8}, and \path{unsloth/Llama-3.2-3B-Instruct}. They are served with vLLM at a 32{,}768-token context cap and a 2{,}048-token output cap, with bf16 weights for Qwen3 8B and Llama, and FP8 weights for Qwen3 14B. Qwen uses temperature/top-p $(0.6,0.95)$ with thinking enabled and $(0.7,0.8)$ with thinking disabled, with top-k~20 and min-p~0 in both modes. Llama uses temperature~0.6 and top-p~0.9. Qwen uses the Hermes tool parser and Qwen3 reasoning parser; Llama uses \texttt{llama3\_json}. The artifact pins the exact model revisions and provides the complete vLLM serving configurations for every supporting run.

The teacher baseline queries \modeldeepseekteacher through OpenRouter with high reasoning effort, a 1{,}000{,}000-token context cap, and a 2{,}048-token output cap. Its evaluated API prices are \$0.14 per million input tokens and \$0.28 per million output tokens. The evaluation evidence released in the artifact repository\textsuperscript{\ref{fn:repo}} includes all primary and supporting static traces.

\begin{table}[!htbp]
  \centering
  \caption{Reward-design results at $r{\le}20$ (Table~\ref{tab:reward_ablation}), with Wilson 95\% CIs and final-checkpoint values. The SFT initialization uses seed 2026 and 300 holdout runs; RL checkpoints use 100 runs. The seed-1337 RQ1 reference uses 500 runs (Table~\ref{tab:e1_trace_design}).}
  \label{tab:e4_reward_ablation_details}
  \footnotesize\setlength{\tabcolsep}{3pt}
  \begin{tabular}{lcrr}
    \toprule
    Reward & Peak & 95\% CI & Final \\
    \midrule
    SFT init. & 76.0\% & 70.9--80.5 & --- \\
    \rewardvariant{Outcome}              & 88.0\% & 80.2--93.0 & 83.0\% \\
    \rewardvariant{{+}Round}        & 90.0\% & 82.6--94.5 & 75.0\% \\
    \rewardvariant{{+}Cost}$^{\star}$       & 90.0\% & 82.6--94.5 & 87.0\% \\
    \rewardvariant{{+}Round{+}Cost}   & 87.0\% & 79.0--92.2 & 79.0\% \\
    \bottomrule
  \end{tabular}
\end{table}

\appendixsubheading{Teacher Inference Configuration}{app:teacher_inference_config}
The supervised trace-collection teacher and the short-reasoning rewriter share a single model and provider route; Table~\ref{tab:teacher_inference_config} reports their request-level settings. All requests use the OpenAI-compatible Chat Completions API and pin the OpenRouter provider to first-party DeepSeek (\path{provider.only=[deepseek]}, \path{provider.allow_fallbacks=false}). Sampling parameters not listed are not set on the request, so trace-collection rollouts sample under DeepSeek-routed OpenRouter defaults.

\begin{table}[!htbp]
  \centering
  \caption{Teacher request-level settings. \textnormal{\modeldeepseekteacher} (\textnormal{\texttt{deepseek/deepseek-v4-flash}}) is accessed through OpenRouter~\cite{openrouter2026pricing} and pinned to the DeepSeek provider.}
  \label{tab:teacher_inference_config}
  \scriptsize\setlength{\tabcolsep}{3pt}
  \begin{tabular}{lccc}
    \toprule
    & Trace (guided) & Trace (unguided) & Short-reason. \\
    & rollout & rollout & rewrite \\
    \midrule
    Temperature    & default\textsuperscript{\textdagger} & default\textsuperscript{\textdagger} & 0 \\
    top-p          & default\textsuperscript{\textdagger} & default\textsuperscript{\textdagger} & default\textsuperscript{\textdagger} \\
    Max output     & default\textsuperscript{\textdagger} & default\textsuperscript{\textdagger} & default\textsuperscript{\textdagger} \\
    Thinking       & disabled & enabled  & disabled \\
    Trace seq.\ len.\ cap & 32{,}768 & 32{,}768 & --- \\
    \bottomrule
    \multicolumn{4}{@{}p{0.96\columnwidth}@{}}{\scriptsize\textsuperscript{\textdagger}Not set on the request; uses DeepSeek provider defaults at trace-collection time.}
  \end{tabular}
\end{table}

\begin{table*}[t]
  \centering
  \caption{Post-hoc prompt-sensitivity audit on the static benchmark. Values are success percentages over $N{=}240$ runs per system-prompt cell. Spread is the range across prompts for each system and budget, computed before rounding.}
  \label{tab:prompt_sensitivity_audit}
  \footnotesize\setlength{\tabcolsep}{6pt}
  \begin{tabular}{lrrrrrrrr}
    \toprule
    & \multicolumn{4}{c}{$r{\le}20$} & \multicolumn{4}{c}{$r{\le}60$} \\
    \cmidrule(lr){2-5}\cmidrule(lr){6-9}
    System & detailed & budget & minimal & spread & detailed & budget & minimal & spread \\
    \midrule
    \modelbackbone & 42.1 & 25.4 & 21.7 & 20.4 & 47.5 & 35.0 & 31.3 & 16.3 \\
    \modelsft      & 80.8 & 76.3 & 75.8 & 5.0  & 82.5 & 79.2 & 80.0 & 3.3 \\
    \modelrl       & 90.0 & 91.7 & 88.3 & 3.3  & 90.0 & 92.1 & 91.7 & 2.1 \\
    \modelgemma    & 80.4 & 82.5 & 77.1 & 5.4  & 89.2 & 86.7 & 83.8 & 5.4 \\
    \bottomrule
  \end{tabular}
\end{table*}

\appendixsubheading{Cost Methodology}{app:cost_methodology}
Local inference cost is estimated as serving-only batched vLLM cost on an RTX~5090 under corporate local-deployment assumptions, including hardware amortization and electricity but excluding agent orchestration, Docker/\allowbreak tool latency, model download, and server cold start. We amortize a \$1{,}999 RTX~5090 plus \$700 host system over 3~years at 50\% utilization (\$0.205/hr), and price 0.675~kW total draw at the Eurostat H2~2025 EU non-household electricity rate of €0.1837/kWh~\cite{eurostatElectricityH22025}, converted at the ECB EUR/USD reference rate to \$0.213/kWh (\$0.144/hr). The resulting hardware-second cost is $C_{\mathrm{sec}}{=}9.70{\times}10^{-5}$~\$/s.

Because vLLM co-schedules prefill and decode under batching, we report a mixed-workload token price calibrated at the throughput/latency knee (concurrency~16):
\begin{equation}
  C_{\mathrm{mix}}^{1\mathrm{M}} =
  C_{\mathrm{sec}}\,
  \frac{T_{\mathrm{bench}}}{N_{\mathrm{in,bench}}+N_{\mathrm{out,bench}}}\cdot 10^6 .
\end{equation}
For \modelbackbone serving this gives $C_{\mathrm{mix}}^{1\mathrm{M}}{=}\$0.0323$/1M tokens. Per-model input/output prices for the local models are fit by replaying real per-round prompts at the same concurrency and anchored to each model's mixed cost, preserving the measured aggregate serving price. Expected cost per successful root at budget $r$ is
\begin{equation}
  C_{\text{root}}(r) =
  \frac{\mathbb{E}[C_{\text{run}}(r)]}{P(H_{\mathrm{root}}\le r)}.
\end{equation}
Under this accounting, \modelrl costs \$0.00213 per successful root at $r{\le}20$. API costs use publicly listed input and output token prices at evaluation time via OpenRouter~\cite{openrouter2026pricing}; \modelclaude costs \$0.1759 per successful root at the same budget. Training-stage cost applies the Verda H100 SXM5 rate of \$2.29 per GPU-hour~\cite{verda2026pricing} to recorded durations on four GPUs: 6{,}006.8683 seconds for SFT and 41{,}578.7137 seconds through RL checkpoint 300, totaling \$121.08. This estimate covers the selected SFT run and RL training through checkpoint 300; it excludes trace collection, ablations, tuning, and evaluation. These prices can change by provider, route, and date, while local costs depend on hardware utilization, batching, \mbox{electricity}, and amortization policy. All reported cost figures are \mbox{therefore setup-specific}.

\section{Interaction-Policy Examples}
\label{app:examples}
To ground the trace-level discussion in the main text, the artifact repository\textsuperscript{\ref{fn:repo}} releases one \modelrl static-benchmark transcript per scenario under \path{examples/}. Each contains the full system prompt, every assistant message, and every tool call and result, with a short header (model, scenario, success flag, turn count). Below, we summarize one success and one Docker failure.

\noindent\textbf{Success, Cron wildcard injection (4 rounds).}
R1 issues 9 parallel recon calls (\texttt{id}, \texttt{sudo -l}, SUID and getcap scans, crontab listing, \texttt{ps}, sudoers) and discovers a root-owned \texttt{/etc/cron.d/backup\_lowpriv}. R2 reads the cron job and writable directories, identifying a \texttt{tar -zcf ... *} pattern over \texttt{/home/lowpriv/backup}. R3 plants the wildcard payload (\texttt{shell.sh}, \path{--checkpoint=1}, \path{--checkpoint-action=exec=sh}\ \texttt{shell.sh}) so the next cron execution sets the SUID bit on \texttt{/bin/bash}. R4 opens an interactive root shell via \texttt{bash -p}.

\noindent\textbf{Failure, Docker group escape (60 rounds, withheld family).}
R1 finds docker-group membership. Subsequent rounds repeatedly launch child containers and probe credentials, reaching root inside containers without obtaining verified root access on the target host. R6 tries to mount the host filesystem (\texttt{-v /:/host}), but the policy never uses \texttt{chroot} to act through that mount. It repeats Docker and credential-check loops until the 60-round budget expires. The failure lies in completing the escape and recovering from ineffective actions, not in detecting Docker access.

\section{Prompt Templates}
\label{app:prompt_templates}
{\raggedright
All prompt templates are released in the artifact repository\textsuperscript{\ref{fn:repo}} under \path{src/prompts/}.
\begin{itemize}
\item \path{privilege_escalation.jinja} is the deployment system prompt shared by \acs{SFT}, \acs{RL}, and static-benchmark evaluation.
\item \path{trace_collection.jinja} adds hidden solution guidance for guided collection, which dataset assembly removes from student inputs.
\item \path{short_reasoning_rewrite.jinja} derives the short-reasoning \acs{SFT} variant.
\item \path{privilege_escalation_budget.jinja} is the \emph{budget} variant of the audit below.
\item \path{privilege_escalation_minimal.jinja} is the \emph{minimal} variant of the same audit.
\end{itemize}
The shared deployment-time user instruction and the no-tool-call recovery nudge stay fixed across all variants and are defined in \path{conf/prompts/default.yaml}.\par}

\appendixsubheading{Prompt-Sensitivity Audit}{app:prompt_sensitivity}
For each model, only the system prompt varies across conditions in Table~\ref{tab:prompt_sensitivity_audit}; tool schemas, the deployment user instruction, the no-tool nudge, and the sampling parameters are held fixed. Here, \emph{detailed} is the canonical deployment prompt used throughout the rest of the paper; \emph{budget} adds an explicit efficiency objective and a short pre-tool reasoning checklist; \emph{minimal} strips procedural guidance to just the task statement, access constraints, and completion condition. Each cell uses 240 new runs, so \emph{detailed} differs slightly from the primary results in Section~\ref{sec:main_results}.

\begin{samepage}
\section*{LLM Usage Statement}
LLMs were used for editorial purposes in this manuscript, and all outputs were inspected by the authors to ensure accuracy and originality. We also used LLM-based coding assistants to help implement the training, evaluation, and analysis code, which the authors reviewed and tested. The research design and interpretation are the authors' own, and they take full responsibility for all content and results.

\acs{LLM}s are also the subject of this research: we study an \acs{LLM} agent trained on traces from an open-weight \acs{LLM} teacher (Section~\ref{sec:agent}), and document all models, prompts, and settings in the appendices. Because the teacher and our models are open-weight and the training and evaluation traces are released, only the \modelclaude baseline depends on a closed model.
\end{samepage}

\end{document}